\documentclass[runningheads]{llncs}
\usepackage{graphicx}
\usepackage{matharticle}
\usepackage[title]{appendix}
\usepackage{comment}
\usepackage{xspace}
\usepackage{booktabs}
\usepackage{paralist}
\usepackage{color}
\usepackage{todonotes}
\newcommand{\fes}{\#\texttt{FEs}\xspace}
\newcommand{\mop}{\textsc{MOP}\xspace}

\newcommand{\saddle}{\textsc{Saddle}}
\newcommand{\nequ}{Nash equilibrium\xspace}

\newcommand{\bneq}{\texttt{BN}\xspace}
\newcommand{\bneqe}{\bneq-\texttt{exact}\xspace}
\newcommand{\bneqa}{\bneq-\texttt{approx}\xspace}

\newcommand{\NE}{\text{NE}\xspace}
\newcommand{\BR}{\texttt{BR}\xspace}
\newcommand{\BO}{BO\xspace}
\newcommand{\GPGp}{\texttt{GPG-psim}\xspace}
\newcommand{\GPGs}{\texttt{GPG-sur}\xspace}
\newcommand{\GPG}{\texttt{GPG}\xspace}

\begin{document}
\title{Approximating Nash Equilibria for Black-Box Games: A Bayesian Optimization Approach\thanks{This material is based upon work supported by the MIT-IBM Watson
		AI Lab and the Defense Advanced Research Projects Agency (DARPA) under Contract No. HR0011-16-C-0101.}}
\titlerunning{Bayesian Optimization for Black-Box Continuous Games}
\author{Abdullah Al-Dujaili\and
Erik Hemberg \and
Una-May O'Reilly}
\authorrunning{A. Al-Dujaili et al.}
\institute{
CSAIL, MIT, USA\\
\email{aldujail@mit.edu, \{hembergerik,unamay\}@csail.mit.edu}}
\maketitle              % typeset the header of the contribution
\begin{abstract}
Game theory has emerged as a powerful framework for modeling a large
range of multi-agent scenarios. Many algorithmic solutions require
discrete, finite games with payoffs that have a closed-form
specification. In contrast, many real-world applications require
modeling with continuous action spaces and black-box utility functions
where payoff information is available only in the form of empirical
(often expensive and/or noisy) observations of strategy profiles. To
the best of our knowledge, few tools exist for solving the class of
expensive, black-box continuous games. In this paper, we develop a
method to find equilibria for such games in a sequential
decision-making framework using Bayesian Optimization. The proposed
approach is validated on a collection of synthetic game problems with
varying degree of noise and action space dimensions. The results
indicate that it is capable of improving the game-theoretic regret in noisy
and high dimensions to a greater extent than hierarchical or discretized
methods.  
\end{abstract}

\section{Introduction}
\label{sec:intro}

Game-theoretic solution concepts play an important role in understanding agents interactions in various areas including ecnonomics~\cite{gibbons1992game}, politics~\cite{morrow1994game}, and cybersecurity~\cite{lye2005game}. The problem of approximating game equilibria and strategic stability has received a lot of attention in the literature. Many of the present solution algorithms and approximation tools are tailored to a restricted class of games. Complex games are then either stylized into a version covered by available solvers (e.g., coarse discretization for infinite games~\cite{kroer2015discretization}) or estimated through simulation and sampling (empirical game-theoretic analysis~\cite{reeves2005generating,wellman2006methods}). Solving finite
approximations to an infinite game can be instructive, but
also produce misleading results~\cite{reeves2004computing}. On the other hand, analyzing games, through simulation and sampling, poses a \emph{search problem} with the goal of identifying equilibrium profiles given a finite number of payoff (utility function) evaluations.  In this paper, our interest is the search problem when such evaluations are expensive and/or noisy. 

We consider one-shot non-cooperative normal-form games, where players make decisions about their actions simultaneously and receive payoffs, upon which the game ends. Our goal is a general-purpose \nequ approximation technique for such games that are expensive, black-box, and continuous.  We report the following contributions: \begin{inparaenum}[1)] \item To the best of our knowledge, this is the first work that approximates equilbria in an expensive, black-box context on the full continuous action space, rather than a discretized representation of the same. \item While majority of equilibria search methods for black-box games cast the problem as the bi-level optimization relying on a best-response search subroutine, we approximate best-responses based on the learned payoff functions. Subsequently, our method solves a flat optimization problem, rather than a hierarchical one.  This saves significant computational expense. \item We validate the effectiveness of our proposition and compare its performance in terms of regret against recent algorithms in the literature on a collection of synthetic games. With about $25$ function evaluations or more, the proposed algorithm is capable of minimizing the regret to an extent which existing tailored methods are not able to reach \item Finally, we provide an implementation for public use. 
\end{inparaenum}

%The paper is structured as follows. Section~\ref{sec:back} presents background and related work. Section~\ref{sec:methods} describes our proposed approach to \nequ approximation. Experiments are reported in Section~\ref{sec:experiments}. Finally, conclusions are drawn and future work is outlined in Section~\ref{sec:conclusion}.
\section{Background}
\label{sec:back}

This section introduces the main terminology---which we adopt from \cite{jordan2008searching,vorobeychik2010probabilistic,picheny2016bayesian}---used in the rest of the paper, followed by a summary of related work.
\subsection{Formal Background}
We start with formalizing the notion of strategic interactions between a set of rational players as follows.
 
\begin{definition}(Normal Form Game~\cite{vorobeychik2008stochastic})
	$(I, \{\calXi\}, \{u_i(\vx) \})$ is a normal form game, with $I$ the set of players where $|I|=p$, $\calXi$ the set of strategies (pure or mixed, depending on context) available to player $i$, $\calX = \prod_{i=1}^{p}\calXi \subseteq \R^{n_\calX}$ the joint strategy set, and $u_i:\calX \to \R$ the utility function for player $i$ mapping the joint strategy $\vx$ to the real-valued payoff received by player $i$ when $\vx$ is played. 
	\label{def:normal-form-game} 
\end{definition}
With \emph{pure strategies}, players choose their \emph{actions} deterministically in the game. Thus, the joint strategy $\vx \in \calX$ represents the joint actions of the set of players $I$, and the term \emph{action} is interchangeable with \emph{strategy}. On the other hand, \emph{mixed strategies} are probability distributions over pure strategies. For brevity, we sometimes omit the word \emph{pure} when referring to a pure (joint) strategy in the rest of the paper. Further, the term \emph{joint strategy}  is interchangeable with \emph{profile}. 

We shall use the convention $\vx = (\vxi, \vxni)$ when the role of player $i$'s action $\vxi$ needs to be emphasized. Accordingly, for player $i$,  we have $u_i(\vx)= u_i(\vxi, \vxni)$. Likewise, $\calX = \calXi \times \calXni$. Holding other players' strategies $\vxni$ constant, player $i$ can best respond by unilaterally deviating from her current strategy $\vxi$ such that her payoff is maximized, as described formally below.

\begin{definition}(Best Response~\cite{jordan2008searching})
	For some joint strategy $\vx \in \calX$, the player $i$'s best-response correspondence is 
	\begin{equation}
	\calB_i(\vx)= \argmax_{\vxi^{\prime}\in \calXi} u_i(\vxi^\prime, \vxni)\;.
	\label{eq:br}
	\end{equation}
	The joint best-response correspondence is then given by
	\begin{equation}
	\calB(\vx)=\prod_{i\in I} \calB_i(\vx)\;.
	\label{eq:joint-br}
	\end{equation}
\end{definition}
The notion of best response is related to another important measure of normal-form games, viz. the \emph{game-theoretic regret} (or simply \emph{regret}). Denoted by $\epsilon(\vx)$, the regret of $\vx$ is the most any player $i$ can gain by deviating from $\vxi$ to any strategy in $\calXi$. Mathematically, we have
\begin{equation}
\epsilon(\vx) = \max_{i \in I} u_i (\calB_i(\vx), \vxni) - u_i(\vx)\;.
\label{eq:regret}
\end{equation} 
Iterative application of best-response correspondence results in the \emph{best-response dynamic}, for which a pure-strategy \emph{\nequ} (\NE) $\vx^*$ is a fixed point. That is, $\vx^*= \calB(\vx^*)$. Nash equilibrium is a game-theoretic solution concept for non-cooperative static games with complete information and no leadership nor followers features~\cite{gibbons1992game}. It formalizes the notion of strategic stability in the sense that every player is playing optimally given other players' choices, as defined next.
\begin{definition} A Nash equilibrium $\vx^* \in \calX$ is the joint strategy  such that
	\begin{equation}
	\forall i \in I\;,\;\; \vxi^* = \calB_i(\vx^*)\;,
	\label{eq:nash-def-1}
	\end{equation}
or equivalently,
\begin{equation}
\epsilon(\vx^*)=0\;.
\label{eq:nash-de-2}
\end{equation}
\end{definition}
A pure \NE does not always exist. Moreover, a game can sometimes be too large to compute a \NE exactly, or only payoff estimates are available. For these cases, $\boldsymbol{\varepsilon}$-\NE is an appropriate solution concept. A profile $\vx\in \calX$ is an $\boldsymbol{\varepsilon}$-\NE if and only if $\epsilon(\vx)\leq \boldsymbol{\varepsilon}$, where $\boldsymbol{\varepsilon}\geq 0$. It follows that every profile $\vx$ is an $\epsilon(\vx)$-\NE, and the pure \NE $\vx^*$ is a $0$-\NE.

In this paper, we are interested in  approximating \textit{pure NEs for continuous black-box games}: \begin{inparaenum}[ i)] \item By \emph{continuous}, we mean that $\{\calXi \subseteq \R^{n_\calXi}\}$. That is, player $i$'s actions are real-valued vectors of size $n_\calXi$. \item By \emph{black-box}, we mean that there is no closed-form expression of the utility functions $\{u_i\}$, or their gradients are neither symbolically nor numerically available. Instead, one  can query an oracle $\voracle$ (e.g., a simulation), which produces a possibly noisy version of $\{u_i\}$ at  specific profile $\vx$. Each oracle call---also referred to as a function evaluation (\texttt{FE})---is often expensive in terms of computational resources (e.g., CPU time). Denoted by $(I, \{\calXi\}, \voracle)$, such games are also referred to as \emph{empirical} or \emph{simulation-based} games~\cite{vorobeychik2008stochastic,vorobeychik2010probabilistic}. Mathematically, we have $\voracle(\vx)=(o_1(\vx),\ldots, o_p(\vx))$ and $\E[\voracle(\vx)]=\vu(\vx)$, where $\vu=(u_1(\vx),\ldots, u_p(\vx))$.
\end{inparaenum} Further, we use the notation $\voracle^{(t)}$ and $o^{(t)}_i$ to denote $\voracle(\vx^{(t)})$ and $o_i(\vx^{(t)})$, respectively. Similar notation will be used for other quantities. In the next section, we summarize literature related to approximating equilibria (if one exists) for expensive, black-box, continuous game-theoretic models.

\subsection{Related Work}

\paragraph{Continuous Games.} For many years, AI researchers have worked on techniques for approximating equilibria in \emph{finite} games (finite action spaces $\calX$). For instance, the \textsf{Gambit}~\cite{mckelvey2016gambit} software package offers a collection of established algorithms to find NEs for finite games. Similarly, theoretical and algorithmic works on \emph{continuous} and \emph{infinite} games are an area of active research. Debreu~\cite{debreu1952social} showed that pure-strategy equilibria exist for infinite games of complete information
with compact, convex action spaces (subsets of a Euclidean space~$\R^{n_\calX}$) and payoffs that are continuous and  quasiconcave in the actions. Fixed-points methods for computing NEs, using standard nonlinear programming routines, were extensively studied~\cite{bacsar1987relaxation,li1987distributed,uryas1994relaxation,krawczyk2000relaxation}. Some propositions employed  coarse discretization amenable to finite-game solvers~\cite{kroer2015discretization}, while others were tailored to restricted classes of games. For  instance, an algorithm was presented in~\cite{reeves2004computing} for two-player games with a class of piecewise linear utility functions. 

\paragraph{Black-Box Continuous Games.} Analytical tools
of game theory have been applied to games that are constructed from empirical observations of strategic play~\cite{wellman2006methods}. The source of these observations is an oracle $\voracle$ representing agent-based simulation or real-world data. The main challenge is to develop algorithms capable of approximating NEs without any information about  the payoffs~$\{u_i\}$, except for a set of $t$ profile-wise observations~$\dset=\{ (\vx^{(1)}, \voracle^{(1)}),\ldots, (\vx^{(t)}, \voracle^{(t)})\}$. Vorobeychik et al.~\cite{vorobeychik2007learning} studied approximating utility functions $\{u_i\}$ given a random sample of pure profiles using supervised learning (regression) methods. Success was measured by how close the players' predicted behavior to that associated with the true utility functions. The $\voracle$-query  complexity (i.e., the number of function evaluations \fes) of
learning equilibria for various classes of games was theoretically addressed in~\cite{fearnley2015learning}. For oracles with noisy outcomes,~\cite{vorobeychik2008stochastic} presented a convergent (in probability) algorithm, which approximates Nash equilibria by minimizing the game-theoretic regret~(\eqref{eq:regret}), based on a hierarchical application of Simulated Annealing (SA) and Monte Carlo methods (to estimate expected payoffs). It was shown empirically that a hybrid method of SA and the Harmony Search algorithm~\cite{geem2010state} converges to an approximate equilibrium point faster than the plain SA at the expense of a slightly lower  approximation accuracy~\cite{alberti2014harmony}.

\paragraph{Expensive Black-Box Continuous Games.} To the best of our knowledge, very little literature addresses game-theoretic models that are expensive to evaluate. The aforementioned hierarchical SA becomes impractical when each oracle call (or simulation) takes nearly an hour~\cite{wellman2005strategic}. At the time of writing this paper, a new version of an article~\cite{picheny2016bayesian} was released on \textsf{arXiv} proposing a novel approach for approximating equilibria in a continuous, black-box, expensive context using Bayesian Optimization (\BO)~\cite{movckus1975bayesian}, which is an established technique to optimize black-box problems. Their proposed approach is realized by fitting Gaussian Processes (GPs) over a coarse discretization of the action space~$\calX$. This is the most relevant work to our proposition and shares the same goal. The main differences are 
\begin{inparaenum}[i)]
	\item \emph{Representation of action spaces $\calX$}: we seek to identify and learn the full game from the limited oracle calls in a way similar to~\cite{vorobeychik2007learning}, while~\cite{picheny2016bayesian} assumes that $\calX$ is either originally discrete, or a representative discretization is available.
	\item \emph{\BO acquisition function}: we employ \BO in minimizing an \emph{approximate} of the game-theoretic regret~(\eqref{eq:regret}) in a way similar to~\cite{vorobeychik2008stochastic}, whereas in~\cite{picheny2016bayesian}, \BO is used to maximize the joint probability of achieving equilibrium or minimizing an uncertainty measure related to equilibrium.
\end{inparaenum}

\section{Methods}
\label{sec:methods}

\begin{figure}
	\centering \includegraphics[width=0.45\textwidth]{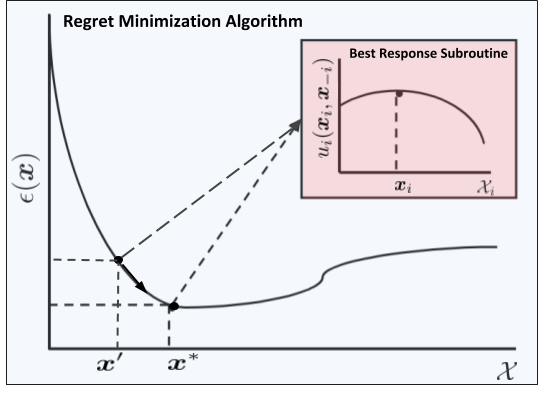}
	\caption{A diagramatic view of finding \nequ based on regret minimization. Adapted from~\cite{vorobeychik2008stochastic}.\label{fig:min-regret}}
\end{figure}

In this section, we describe our proposition, which we refer to  as \bneq: \underline{\texttt{B}}ayesian optimization to approximate \underline{\texttt{N}}ash equilibria given a finite number of oracle $\voracle$ queries. The approach we take is similar to~\cite{vorobeychik2008stochastic}: to minimize the regret~(\eqref{eq:regret}). As shown in Figure~\ref{fig:min-regret}, regret-minimization approaches usually employ best-response $\{\calB_i(\vx)\}$ approximation as a subroutine to estimate the regret $\epsilon(\vx)$ at a given profile $\vx \in \calX$. This entails solving a bi-level optimization problem, which can require a prohibitive number of function (oracle) evaluations. This poses a challenge in expensive settings.
Instead of using best-response approximation as a subroutine~\cite{vorobeychik2008stochastic}, we make use of GPs to jointly learn the payoffs and approximate the regret based on $\dset$: the set of empirical observations of strategic plays obtained after $t$~ function evaluations (\texttt{FEs}). At the same time, we employ the \BO  framework to search for a NE $\vx^*$ by sampling the profile space~$\calX$ sequentially. The sequential sampling is guided by optimizing the approximated regret $\hat{\epsilon}(\cdot|\dset)$, which represents the so-called \emph{acquisition function} in \BO literature. Optimizing~$\hat{\epsilon}$ can be carried out by off-the-shelf global optimization algorithms, as it does not make any call to the oracle $\voracle$ and hence it is relatively inexpensive to evaluate. Mathematically, at step $t\geq t_o$,
$$\vx^{(t+1)}\in \argmin_{\vx\in \calX} \hat{\epsilon}(\vx|\dset).$$
The profile $\vx^{(t+1)}$, which is a potential NE, represents the next strategic play to be evaluated by invoking the oracle~$\voracle$. At step $t+1$, the GP models are fitted with 
$$\mathcal{D}^{1:t+1}=\dset \cup \{(\vx^{(t+1)}, \voracle(\vx^{(t+1)})) \}\;,$$
and the procedure continues iteratively. After $T$ \texttt{FEs}, the algorithm returns the profile  $\vx(T)$ of the lowest obtained regret from the constructed sequence $\{\vx^{(1)},\ldots, \vx^{(T)} \}$. Note that,  $\voracle(\vx)$ corresponds to a single (possibly noisy) sample of the payoffs value at profile $\vx$, in contrast to the Monte Carlo estimate (averaging several samples) employed in the hierarchical SA approach of~\cite{vorobeychik2008stochastic}. We consider models with additive noise corruption, that is
\begin{equation}
o_i(\vx) = u_i(\vx) + \zeta_i\;,\; \forall i \in I,
\end{equation}
where $\zeta_i \sim \normal(0, v_i)$ with $v_i\geq 0$. 

A GP is a distribution over functions
specified by its mean $\mu(\cdot)$ and covariance (or kernel)
$k(\cdot, \cdot)$ functions. Given the profiles $(\vx^{(1)}, \ldots, \vx^{(t)})$, denoted by $\vx^{1:t}$, and the corresponding observed player $i$'s payoffs $(o^{(1)}_i, \ldots, o^{(t)}_i)$, denoted by $o^{1:t}_i$, we have
\begin{equation}
o^{1:t}_i \sim \normal \big(\mu^{1:t}_i , \vKi\big)\;,\; \forall i \in I\;,
\label{eq:gp}
\end{equation}
where $\mu^{1:t}_i=(\mu^{(1)}_i, \ldots, \mu^{(t)}_i)$ and $\vKi$ is the covariance matrix, with its entries~$\vKi^{(p,q)} = k_i(\vx^{(p)}, \vx^{(q)})$. Common choices of $\{k_i(\cdot, \cdot)\}$ in \BO literature are the squared exponential and Mat\'{e}rn kernels. For more details on GPs, we refer the reader to~\cite{rasmussen2004gaussian}. Our choice of GPs was primarily due to their analytical tractability, which enable us to compute exactly the posterior predictive mean $\mu_i(\cdot|\dset)$ and variance~$\sigma^2(\cdot|\dset)$ for any profile $\vx \in \calX$, as well as to approximate the regret $\epsilon(\vx)$~(\eqref{eq:regret}) by~$\hat{\epsilon}(\cdot|\dset)$. Recall that computing the regret requires the values $u_i(\vxi,\vxni)$ and $u_i(\calB(\vx), \vxni)$ for all $i\in I$. In the literature~\cite{vorobeychik2008stochastic,vorobeychik2010probabilistic}, the former quantity has been estimated by invoking the oracle multiple times for $\vx$ and averaging the obtained observations, while the latter quantity is estimated by a stochastic search method (e.g., SA), which also invokes the oracle multiple times at each point of its search trajectory. Here, we approximate $u_i(\vxi,\vxni)$ and $u_i(\calB(\vx), \vxni)$ with GPs.

\paragraph{Approximating $u_i(\vxi,\vxni)$.} Given $\dset$ and  assuming that the prior mean function $\mu_i(\cdot)=0$, the posterior predictive distribution of $o_i(\vx^{(t+1)})$ is of the form~\cite{rasmussen2004gaussian}
\begin{equation}
o^{(t+1)}_i | \dset \sim \normal(\mu_i(\vx^{(t+1)}|\dset), \sigma^2(\vx^{(t+1)}|\dset))\;, \forall i \in I\;,
\label{eq:predictive-distribution}
\end{equation}
where
\begin{align}
\mu_i(\vx^{(t+1)}|\dset)&=\vki(\vx^{(t+1)})^{T}\vKi^{-1}o^{1:t}_i\;, \nonumber \\
\sigma^2_i(\vx^{(t+1)}|\dset)&=k_i(\vx^{(t+1)},\vx^{(t+1)}) - 
\vki(\vx^{(t+1)})^{T}\vKi^{-1}\vki(\vx^{(t+1)})\;, \nonumber\\
\intertext{with} 
\vki(\vx^{(t+1)})&= (k_i(\vx^{(t+1)},\vx^{(1)}) , \ldots, k_i(\vx^{(t+1)},\vx^{(t)}) )\;, \nonumber\\
k_i(\vx, \vx^{\prime})&= v_i \delta(\vx-\vx^{\prime}) +  c_i\exp\bigg(-\frac{(\vx-\vx^{\prime})^TD_i (\vx-\vx^{\prime})}{2}\bigg) \nonumber \;.
\end{align}
One can observe that the kernel function $k_i(\cdot, \cdot)$ is a combination of the white kernel  and the scaled squared exponential kernel to explain the noise-component of the observations. The kernel's hyperparameters $v_i$, $c_i$, and the diagonal matrix $D_i$ are tuned by maximizing the GP log-marginal likelihood.

As mentiond in Section~\ref{sec:back}, $\E[\voracle(\vx)]=\vu(\vx)$. Thus, according to our GP model, for all $i\in I$, $u_i(\vxi,\vxni)$ is approximated by
\begin{equation}
u_i(\vxi,\vxni)\approx \E[o_i(\vx) | \dset]=\mu_i(\vx|\dset)\label{eq:mu}\;.
\end{equation}
Next, we show how GPs can be used to approximate $u_i(\calB(\vx), \vxni)$.

\paragraph{Approximating $u_i(\calB(\vx), \vxni)$.} This value corresponds to the maximum of the set of values $$\uivxni\defeq\{ u_i(\vxi^\prime, \vxni) \mid \vxi^\prime \in \calXi \} \subset \R\;.$$ 
Assuming these values are finite (bounded), one may recover its maximum using its mean and standard deviation, as shown in the motivating example below.

\begin{example} Let $\uivxni$ be the support of a uniform distribution $\mathcal{U}([a,b])$, then the maximum of this set, which is $b$ and in our setup it is $u_i(\calB(\vx), \vxni)$, can be recovered from the distribution mean ${(a+b)}/{2}$ and standard deviation $(b-a)/\sqrt{12}$ as $$\frac{a+b}{2} + \gamma \frac{b-a}{\sqrt{12}}$$ where $\gamma=\sqrt{3}$.
\end{example}
Similarly, we approximate $u_i(\calB(\vx), \vxni)$, the payoff of $i$'s best response to $\vxni$, according to our GP models  as
$$u_i(\calB(\vx), \vxni) \approx \bar{\mu}_i(\vx|\dset) + \gamma \bar{\sigma}_i(\vx | \dset)\;,$$
where $\bar{\mu}_i(\vx|\dset)$ and $\bar{\sigma}_i(\vx | \dset)$ are the mean and standard deviation of $\uivxni$, respectively. Formally, we have
\begin{eqnarray}
\bar{\mu}_i(\vx|\dset)&=& \E_{\vxi^\prime}[\mu_i(\vxi^\prime, \vxni|\dset)]\;,
\label{eq:barmu} \\
\bar{\sigma}^2_i(\vx | \dset)&=& {\E_{\vxi^\prime}[(\mu_i(\vxi^\prime, \vxni|\dset)- \bar{\mu}_i(\vx|\dset))^2]}\;,
\label{eq:barstd}
\end{eqnarray}
and $\gamma\geq 0$ is a hyperparameter of our algorithm. An appropriate value can be the $99^{th}$ percentile of the standard normal distribution $\approx2.33$. Provided that the covariance functions $\{k_i\}$ of our GP models are separable, the values of~\eqref{eq:barmu} and~\eqref{eq:barstd} can be computed exactly (in a closed-form), as shown in Sections~\ref{sec:mu-proof} and~\ref{sec:std-proof} of the supplement materials.

\paragraph{Approximating the regret~(\eqref{eq:regret}).} Based on the above, in particular~\eqref{eq:mu},~\eqref{eq:barmu}, and~\eqref{eq:barstd}, the game-theoretic regret $\epsilon(\vx)$ can be estimated as
\begin{equation}
\hat{\epsilon}(\vx|\dset)= \max_i \bar{\mu}_i(\vx|\dset) + \gamma \bar{\sigma}_i(\vx | \dset)  -  \mu_i(\vx|\dset)\;.
\label{eq:regret-approx}
\end{equation}
This constitutes the \emph{acquisition function} of our \BO framework. In our implementation, we rescale this term by $\bar{\sigma}_i(\vx | \dset)$ to have similar sensitivity across $\vx\in \calX$. This improves the performance of the global optimization procedure in its search for the next play $\vx^{(t+1)}$.

\paragraph{Exploration-Exploitation Trade-off.} In general, the acquisition function in the \BO framework balances between exploration (sampling from areas in $\calX$ of high uncertainty according to the GP model) and exploitation (sampling from potentially optimal areas in $\calX$ according to the GP model). One could observe that the acquisition function represented by the estimated regret~(\eqref{eq:regret-approx}) is more exploitative than exploratory. This can be addressed with several policies from the literature~\cite{auer2002finite}. We employ an $\varepsilon$-greedy policy with $\varepsilon\in (0,1)$,\footnote{This $\varepsilon$ is different from the $\boldsymbol{\varepsilon}$ in $\boldsymbol{\varepsilon}$-NE. The two quantities happen to have the same symbols. We sought to differentiate them with a bold version in the case of $\boldsymbol{\varepsilon}$-NE.} where the next strategic play $\vx^{(t+1)}$ is chosen with probability $1-\varepsilon$ according to the approximated regret $\hat{\epsilon}(\cdot|\dset)$. Otherwise, it is chosen according to the posterior uncertainty~$\sigma_i(\cdot|\dset)$.

\begin{figure}[h!]
	\centering
	\begin{tabular}{cc}
		\includegraphics[width=0.4\textwidth]{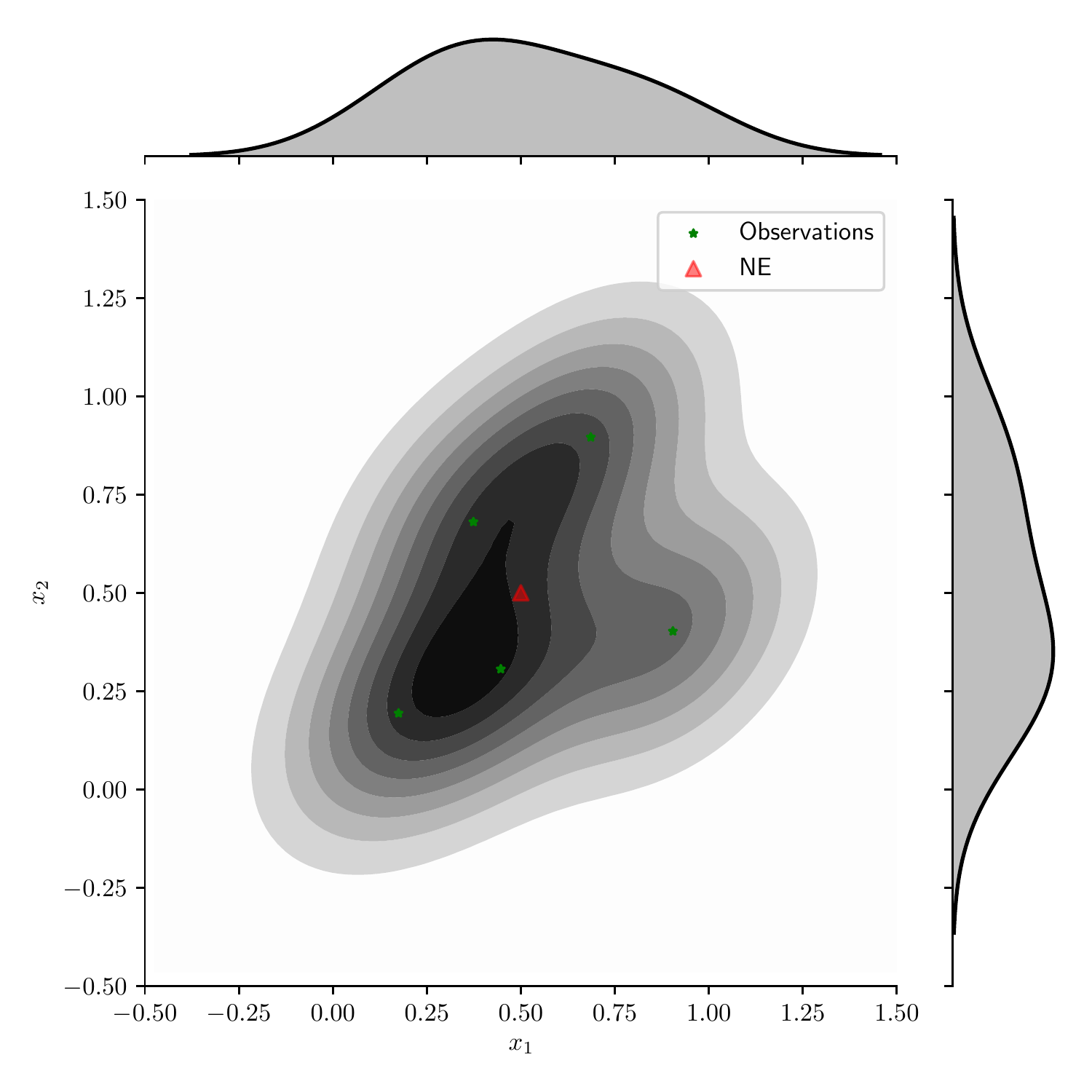} &
		\includegraphics[width=0.4\textwidth]{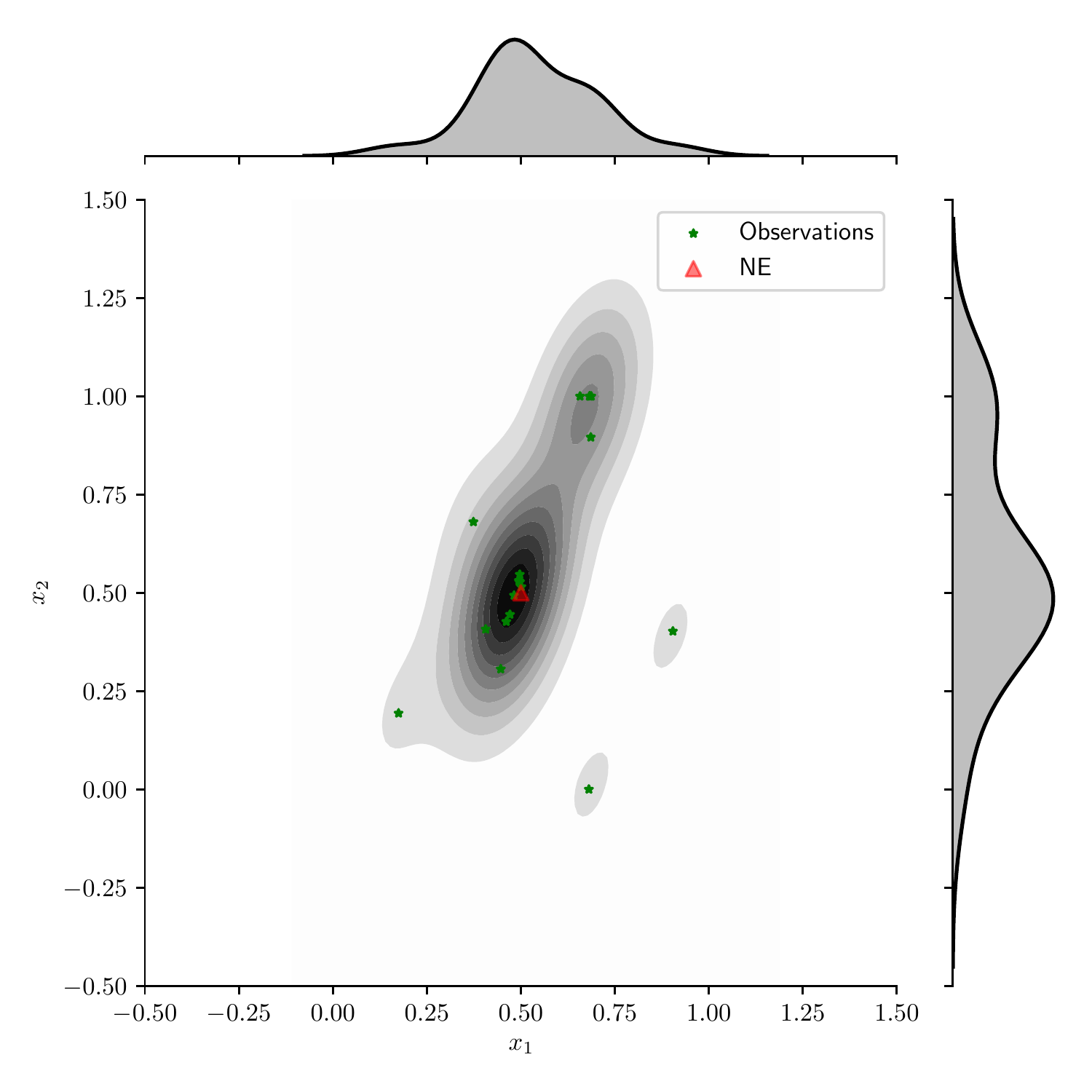} \\
		\tiny (a) Sampled profiles after $5$ observations& \tiny (b) Sampled profiles after $20$ observations\\
		\includegraphics[width=0.42\textwidth,clip,trim=62 100 50 100]{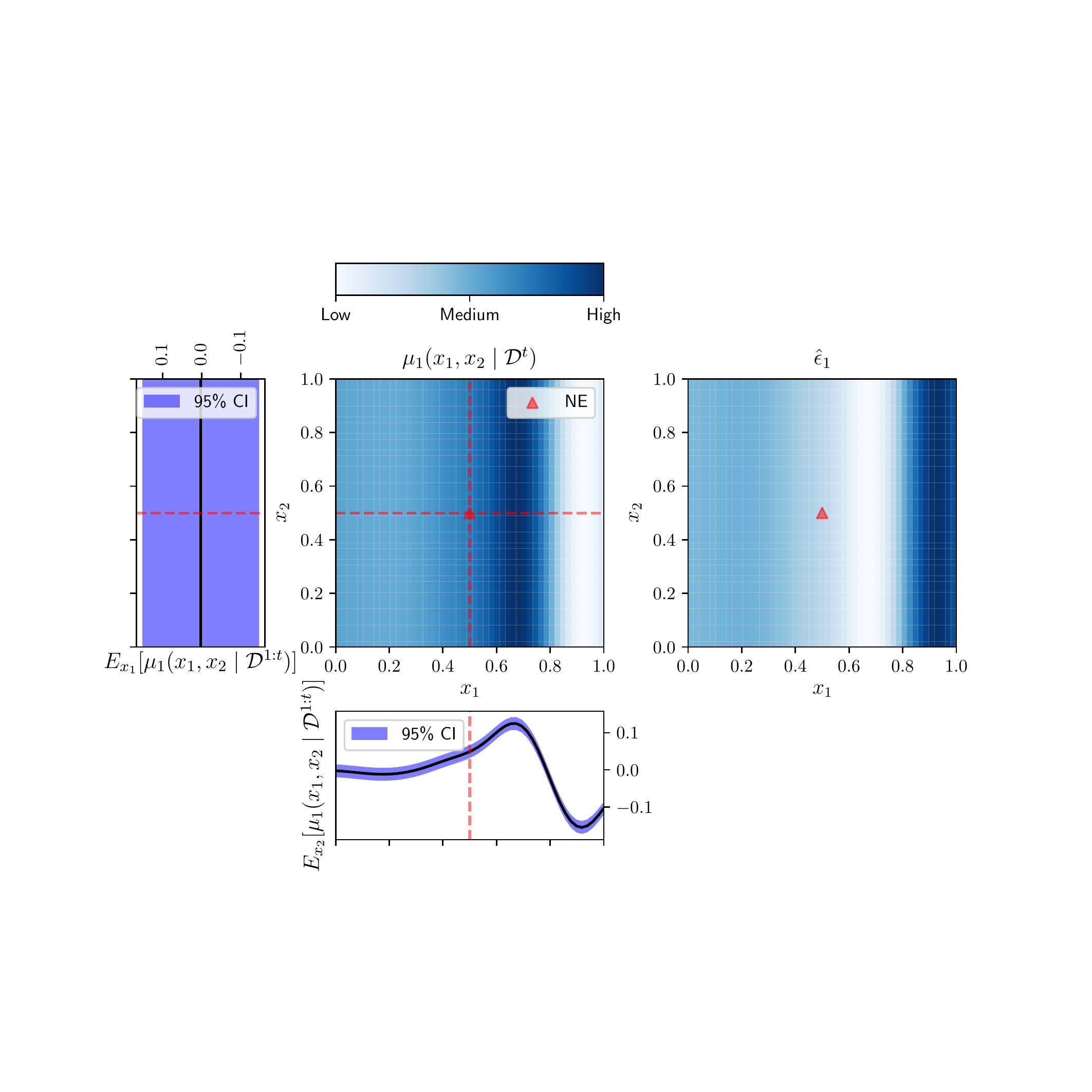} &
		\includegraphics[width=0.42\textwidth,clip,trim=50 100 62 100]{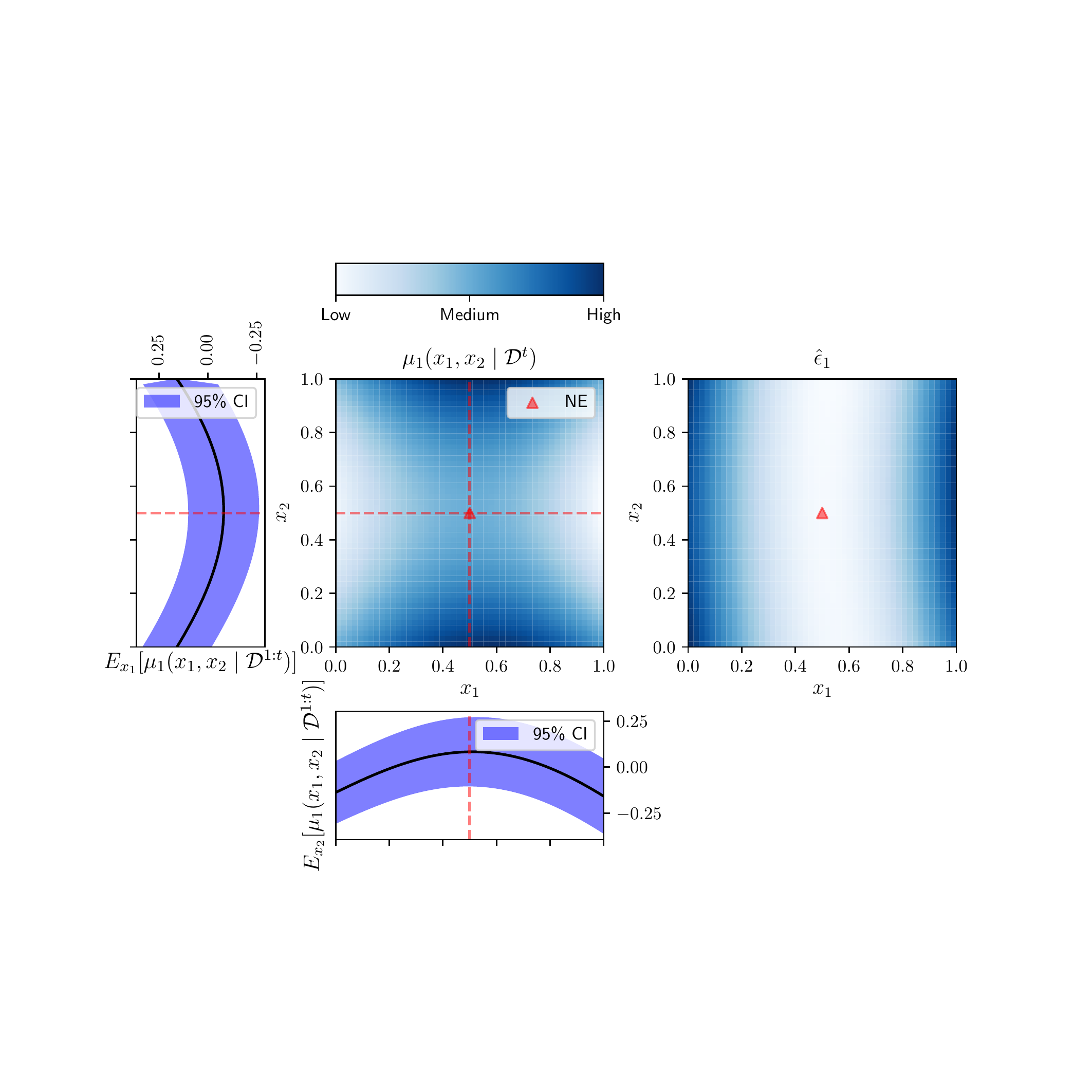} \\
		\tiny (c) Player $1$ after $5$ observations &
		\tiny (d) Player $1$ after $20$ observations \\
		\includegraphics[width=0.42\textwidth,clip,trim=62 100 50 100]{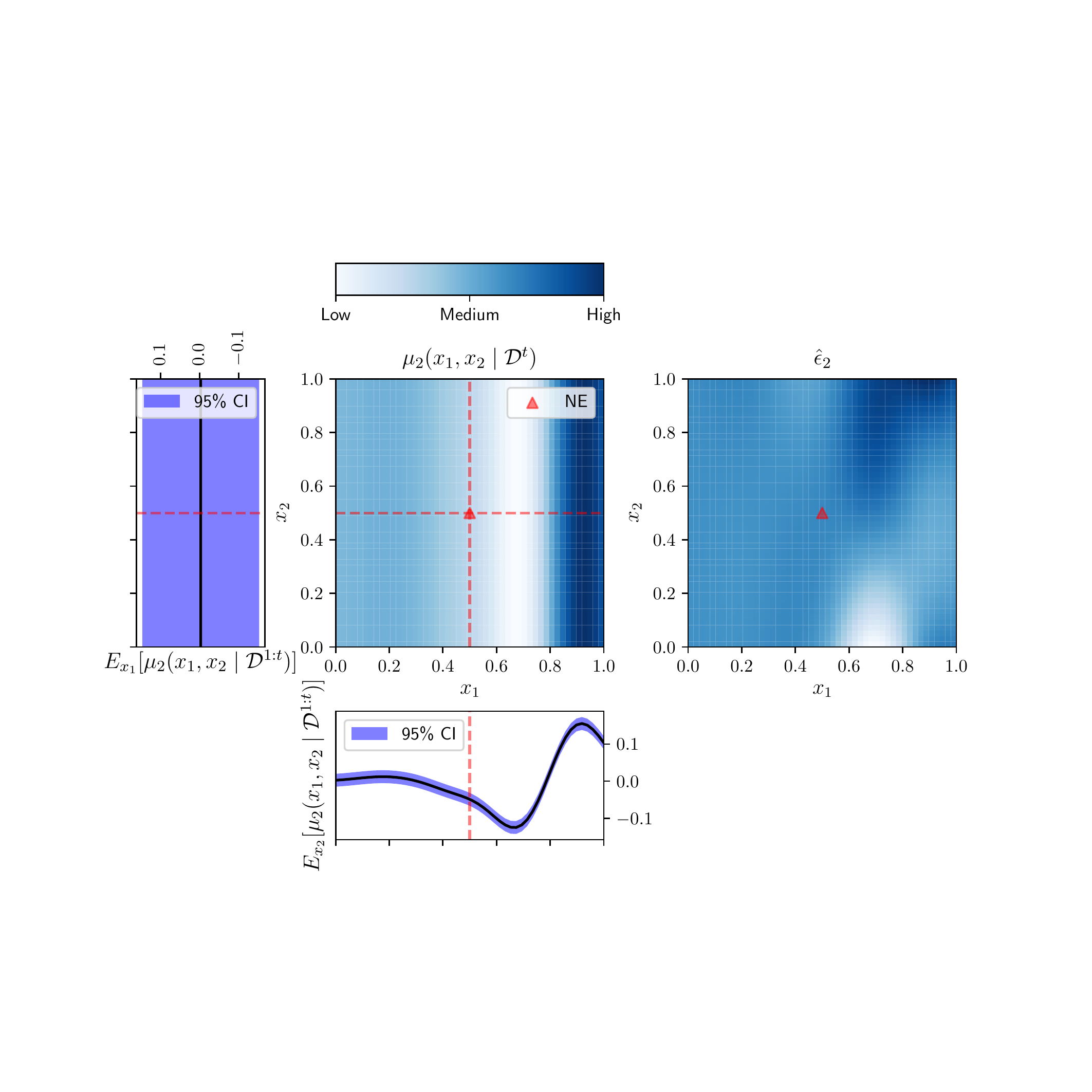} &
		\includegraphics[width=0.42\textwidth,clip,trim=50 100 62 100]{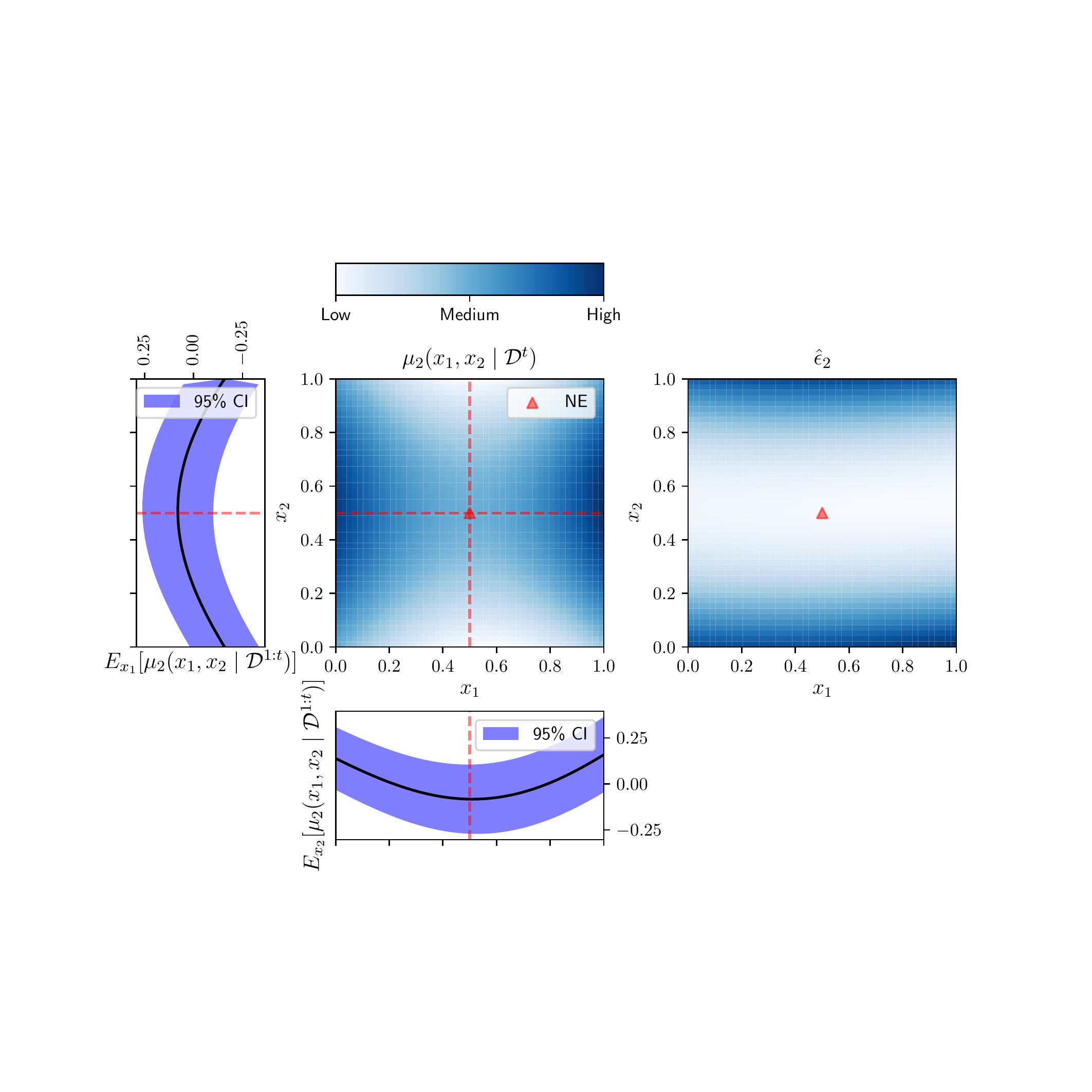} \\
		\tiny (e) Player $2$  after $5$ observations & \tiny
		(f) Player $2$ after $20$ observations \\
	\end{tabular}
	\caption{Illustration of \bneqe after $5$ and $20$ observations on the zero-sum hyperbolic paraboloid game with NE being $(0.5,0.5)$. That is, $u_1(x_1, x_2)= (x_2 - 0.5)^2 - (x_1 - 0.5)^2$, and $u_2(x_1, x_2)=-u_1(x_1, x_2)$. $\hat{\epsilon}_i$ denotes $(\bar{\mu}_i(\vx|\dset) + \gamma \bar{\sigma}_i(\vx | \dset)  -  \mu_i(\vx|\dset))/ \bar{\sigma}_i(\vx | \dset)$, where $\gamma=2.32635$, the $99^{th}$ percentile of the standard normal distribution.}
	\label{fig:saddle-illust-exact}
\end{figure}

\paragraph{Computational Tractability.} With the use of separable kernels, we could compute $ \bar{\mu}_i(\vx|\dset) $ and $\bar{\sigma}_i(\vx | \dset)$ exactly and evaluate the approximated regret~$\hat{\epsilon(\vx|\dset)}$. However, the computational complexity, besides the cost of computing and inverting the covariance matrix $K_i$,  of computing the regret grows quadratically in the number of observations $|\dset|$. This could be manageable in an expensive setup where the number observations is well below 100, i.e.,  $|\dset| \leq 100$. Beyond that, one could estimate these values through sampling. Mathematically, for all $i\in I$,
\begin{eqnarray}
\hat{\bar{\mu}}_i(\vx|\dset)&=&\frac{1}{S}\sum^{S}_{s=1}\mu_i(\hat{\vx}_i^{(s)}, \vxni|\dset)\;, \label{eq:bbarmu} \\
\hat{\bar{\sigma}}^2_i(\vx | \dset)&=&\frac{1}{S}\sum^{S}_{s=1} (\mu_i(\hat{\vx}_i^{(s)}, \vxni|\dset) - \hat{\bar{\mu}}_i(\vx|\dset))^2\;,\label{eq:bbarstd}
\end{eqnarray}
where $S$ is the number of samples, and $\hat{\vx}_i^{(s)}\sim \calU(\calXi)$ or any relevant sampling technique. This approximation can also be used for GPs with inseparable kernels. Note that there exist other methods to approximate the integrals associated with the inseparable kernel~\cite{briol2015frank}. On the other hand, the noise and randomness introduced by sampling may help towards a better exploration-exploitation trade-off. Our implementation supports both exact and approximate computation of $ \bar{\mu}_i(\vx|\dset) $ and $\bar{\sigma}_i(\vx | \dset)$ and we refer to these algorithmic variants by \bneqe and \bneqa, respectively. With this at hand, our general-purpose framework for computing Nash equilibria in expensive, black-box, continuous games is now complete. Algorithm~\ref{alg:bneq} summarizes the procedure and we provide an illustration of the same (the exact variant~\bneqe) in Fig.~\ref{fig:saddle-illust-exact}. 
The figure shows the progress of \bneq variants from $5$ to $20$ observations. The top subfigures (a and b) show the sampled profiles and their kernel density estimate (KDE) over the action space~$\calX$. With more iterations, the sampled profiles get closer to the actual NE. The middle subfigures (c and d) show the estimated payoff and approximated regret for player~$1$ after $5$ and $20$ observations, respectively. Likewise, the bottom subfgures (e and f) show the same for player~2. Since subfigures (c), (d), (e), and (f) are similar. We describe them collectively. The center heatmap represents the GP's current estimate of player~$i$'s payoff $\mu_i(x_1, x_2|\calD^{1:t})$~(\eqref{eq:mu}). The supplots to the left and bottom show the expectation (along with a 95\%-confidence interval) of $\mu_i(x_1, x_2|\calD^{1:t})$ over $x_1$ and $x_2$, respectively. For player $1$, the \emph{left} supplot of subfigures (c) and (d)  
represent ${\bar{\mu}}_1(\vx|\calD^{1:5})$~(\eqref{eq:barmu}) and ${\bar{\mu}}_1(\vx|\calD^{1:20})$ in Fig.~\ref{fig:saddle-illust-exact}. For player $2$, the \emph{bottom} subplots in subfigures (e) and (f) correspond to ${\bar{\mu}}_2(\vx|\calD^{1:5})$ and ${\bar{\mu}}_2(\vx|\calD^{1:20})$ in Fig.~\ref{fig:saddle-illust-exact}. For player~$1$ (i.e., subfigures (c) and (d)), this is computed based on the center and the left subplots. For player~$2$ (i.e., subfigures (e) and (f)), this is computed based on the center and the \emph{right} subplots. One can observe the smoothness of the plots in the figures of \bneqe (Fig.~\ref{fig:saddle-illust-exact}). For further illustrations, we refer the reader to Section~\ref{sec:fig-illust} of the supplement materials.

\begin{algorithm}
	\small
	\caption{Bayesian optimization for Nash Equilibrium (\bneq) \newline
		\textbf{Require:} \newline 
		$~~$\hspace{\algorithmicindent}$\calD^{1:t_0}$: initial design  (e.g., Latin hypercube design~\cite{mckay2000comparison})\newline
		$~~$\hspace{\algorithmicindent}$T$: number of iterations\newline 
		$~~$\hspace{\algorithmicindent}$\varepsilon\in (0,1)$: probability of exploration
	}
	\label{alg:bneq}
	\begin{algorithmic}[1]
		\For{$t=1$~\textbf{to}~$T$}		
		\State \label{ln:acq-fct}$\vx^{(t+1)} \gets 
				\begin{cases}
				 \argmin_{\vx\in \calX} \hat{\epsilon}(\vx|\dset)\;, & \text{with probability $1-\varepsilon$;} \\
				 \argmax_{\vx\in \calX} \max_{i\in I} \sigma_i(\vx|\dset)\;, & \text{otherwise.}
				\end{cases}$
			   
		\State $\calD^{1:t+1} \gets \dset \cup \{(\vx^{(t+1)}, \voracle(\vx^{(t+1)})\}$
		\EndFor
	\end{algorithmic}
\end{algorithm}

While the BO framework provides an efficient sampling of the profile space~$\calX$, one should note that deciding which point (from $\calD^{1:T}$) corresponds to the approximate NE, $\vx(T)$, can be prone to over-fitting~\cite{cawley2007preventing}. This is not a problem in the standard BO where the best solution $\vx(T)$ can be decided directly by comparing the values returned by the oracle. This is beyond the scope of this paper and we leave it for future work.
\section{Experiments}
\label{sec:experiments}

In this section, we investigate how \bneq performs and compare it
with~\cite{picheny2016bayesian}'s Gaussian process method (\GPG), which discretizes $\calX$, and the iterated best-response (\BR) via stochastic search~\cite{vorobeychik2008stochastic}. We are interested in the impact on performance of
using a continuous representation, in contrast to the grid that \GPG
uses. In addition, we are interested in how each algorithm scales and is robust to a noisy payoff function. Our experimental problems are shown in Table~\ref{tab:problemSetup}, the columns
show the action space for each player and the given number of function evaluations (\fes). \saddle.$\{1, 2,3\}$ are variants of the zero-sum hyperbolic paraboloid game where the \NE $\vx^*$ is shifted or the game is scaled to a higher-dimensional action space. That is, $u_1(x_1, x_2)= (x_2 - x^*_2)^2 - (x_1 - x^*_1)^2$, and $u_2(x_1, x_2)=-u_1(x_1, x_2)$. The \mop problem is taken
from~\cite{picheny2016bayesian}. In line with~\cite{dewancker2016stratified}, all
algorithms were terminated after $20n_\calX$~\texttt{FEs}. For comparison with \GPG which evaluates at every discretized grid point, with \saddle.3 we terminated the algorithms after \fes equal to grid size ($120$). For experiments with noise, a Gaussian noise was added to the payoff functions with a standard deviation $\sigma^\prime_i$ that is roughly $0.1$ of its range: For \mop, $\sigma^\prime_1=7.5$ and $\sigma^\prime_2=3$. For \saddle~variants, $\sigma^\prime_1=\sigma^\prime_2=0.025$.

\begin{table}
	\centering
	\caption{Problem setup. The columns show the dimensions for
          each player and the number of fitness evaluations.}
        \label{tab:problemSetup}
     \resizebox{0.75\textwidth}{!}{
	\begin{tabular}{p{1.75cm}p{3cm}p{3cm}p{4cm}}
		\toprule
		\textbf{Problem}  & \boldsymbol{$(\calX_1, \calX_2)$} & \textbf{NE} & Function evaluations (\fes)\\ \toprule
		\saddle.1 &  $([0,1],[0,1])$ &  $(0.5, 0.5)$ & 40 \\ \midrule
		\saddle.2 & $([0,1],[0,1])$ & $(0.3, 0.3)$ & 40 \\ \midrule
		\saddle.3 & $([0,1]^2,[0,1]^2)$ & $([0.5]^{2}, [0.5]^{2})$ & 120\\ \midrule
		\mop & $([0,1],[0,1])$ & $(0.08093,1)$ & 40\\ \midrule
	\end{tabular}
}
\end{table}
\paragraph{\bneq setup.} The source for \bneq and the supplement materials are available at \url{https://github.com/ALFA-group}. The algorithm is implemented  in \textsf{Python} and uses the~\textsf{Scikit-learn} package~\cite{scikitlearn}. The hyperparameters that
are not default in \bneq were set to:
\begin{inparaenum}[A)]
\item The bounds of kernel $\{k_i\}$'s hyperparameters  were set as follows.
	$v_i \in [10^{-5}, 10^{5}]$, $c_i \in [10^{-3}, 10^{3}]$, and  $D^{l,l}_i \in [10^{-2}, 10^{2}]$. The hyperparameters were tuned with $2$ restarts.
\item The size of the initial design $|\dset|$ is set to $\lfloor \fes / 4 \rfloor$.
\item The acquisition function (Line~\ref{ln:acq-fct} of Algorithm~\ref{alg:bneq}) is optimized with \texttt{CMA-ES}~\cite{hansen2003reducing} and an evaluation budget of $250$.
\item The $\varepsilon$-greedy policy is set with $\varepsilon=0.05$.
\item For \bneqa, $10n_{\calXi}$ samples from the Latin hypercube were used to compute $\hat{\bar{\mu}}_i(\vx|\dset)$~(\eqref{eq:barmu}) and $\hat{\bar{\sigma}}_i(\vx | \dset)$~(\eqref{eq:barstd}).
\end{inparaenum}
\paragraph{\GPG setup.} We tested two variants of the algorithm: \GPGp and \GPGs with \texttt{"psim"} and \texttt{"sur"} as the solver criterion, respectively. Similar to \bneq, $\lfloor \fes / 4 \rfloor$ are used as initial points. We used a $31 \times
31$ grid for all the problems, except for \saddle.3 that has $11 \times 11$. The rest of the parameters were set to their default settings.

\paragraph{\BR setup.} In our \BR implementation, we used \texttt{basinhopping} from the \texttt{SciPy} package~\cite{scipy}, with \texttt{L-BFGS-B} as the local minimization method. \BR is prohibitively expensive on problems with noisy observations, as it requires multiple samples per profile. Therefore, \BR is only tested on noiseless experiments for baseline comparison to the methods tailored for expensive games.

\begin{figure}[t!]
	\begin{tabular}{ccc}
		\tiny (a)  Noiseless \saddle1~&
		\tiny (c)  Noiseless \saddle2 &
		\tiny (e)  Noiseless \mop \\
		\includegraphics[width=0.33\textwidth]{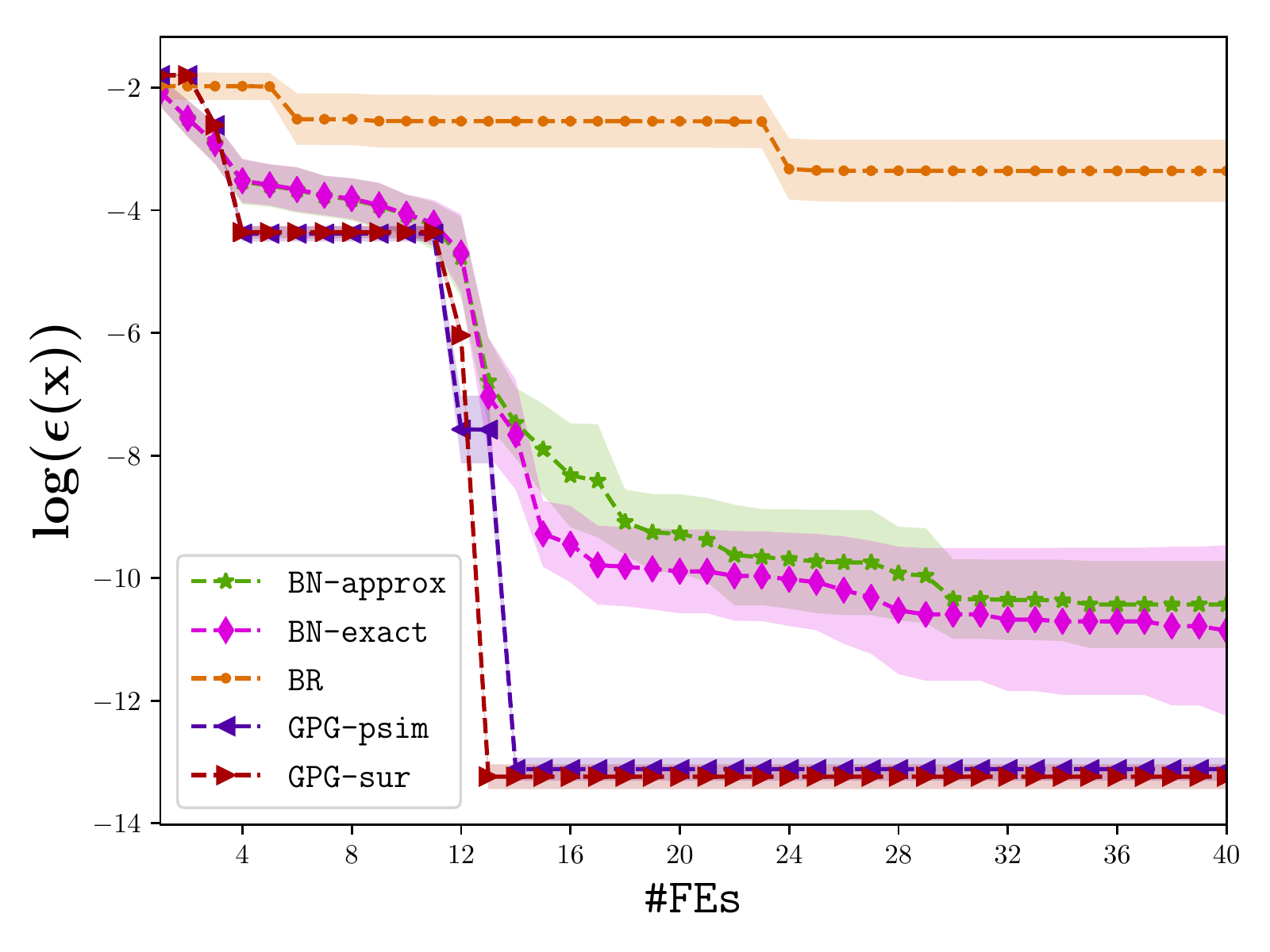} & 
		\includegraphics[width=0.33\textwidth]{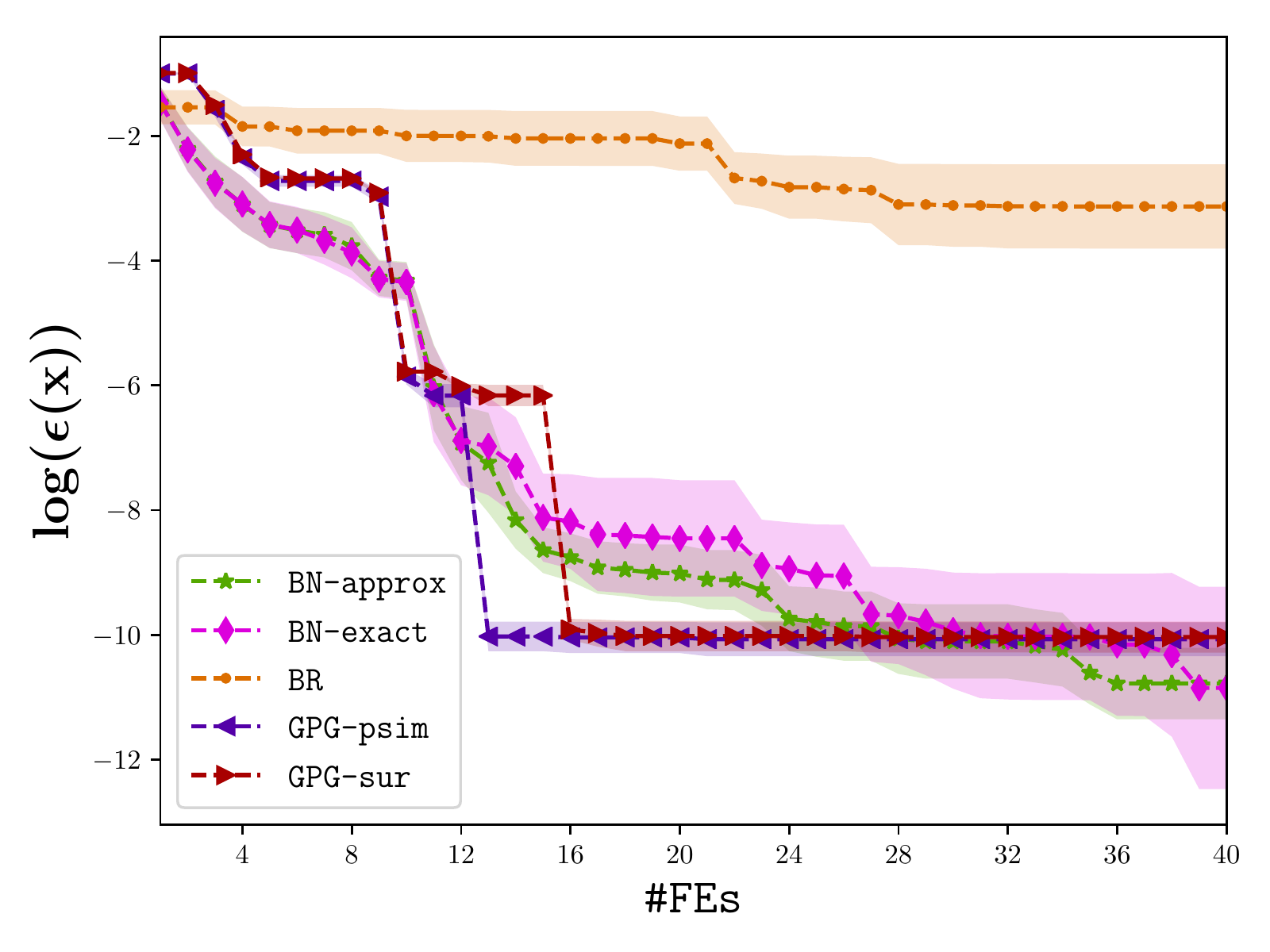} &
		\includegraphics[width=0.33\textwidth]{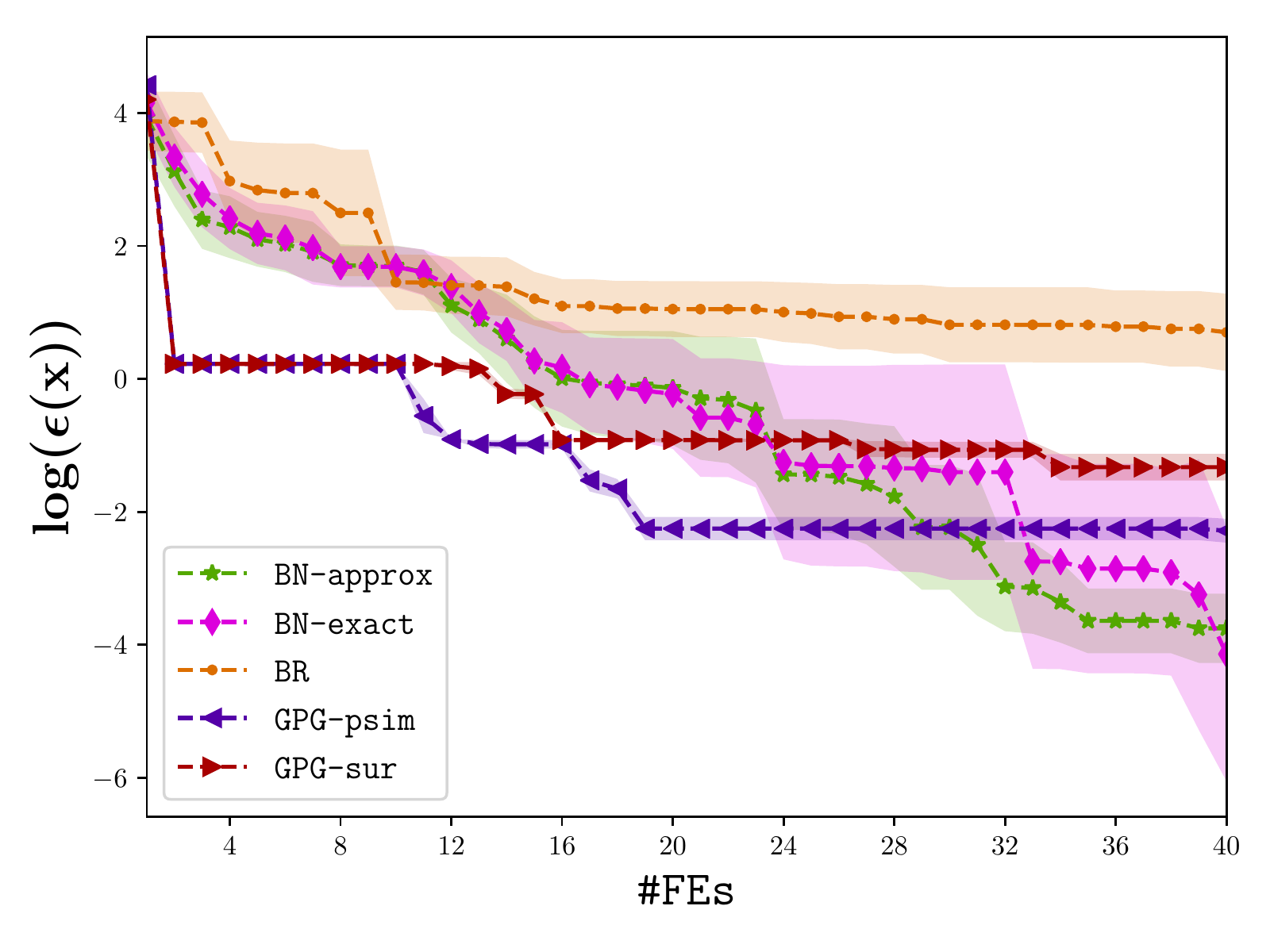} \\
		\tiny (b)  Noisy \saddle1~&
		\tiny (d)  Noisy \saddle2~&
		\tiny (f)  Noisy \mop\\
		\includegraphics[width=0.33\textwidth]{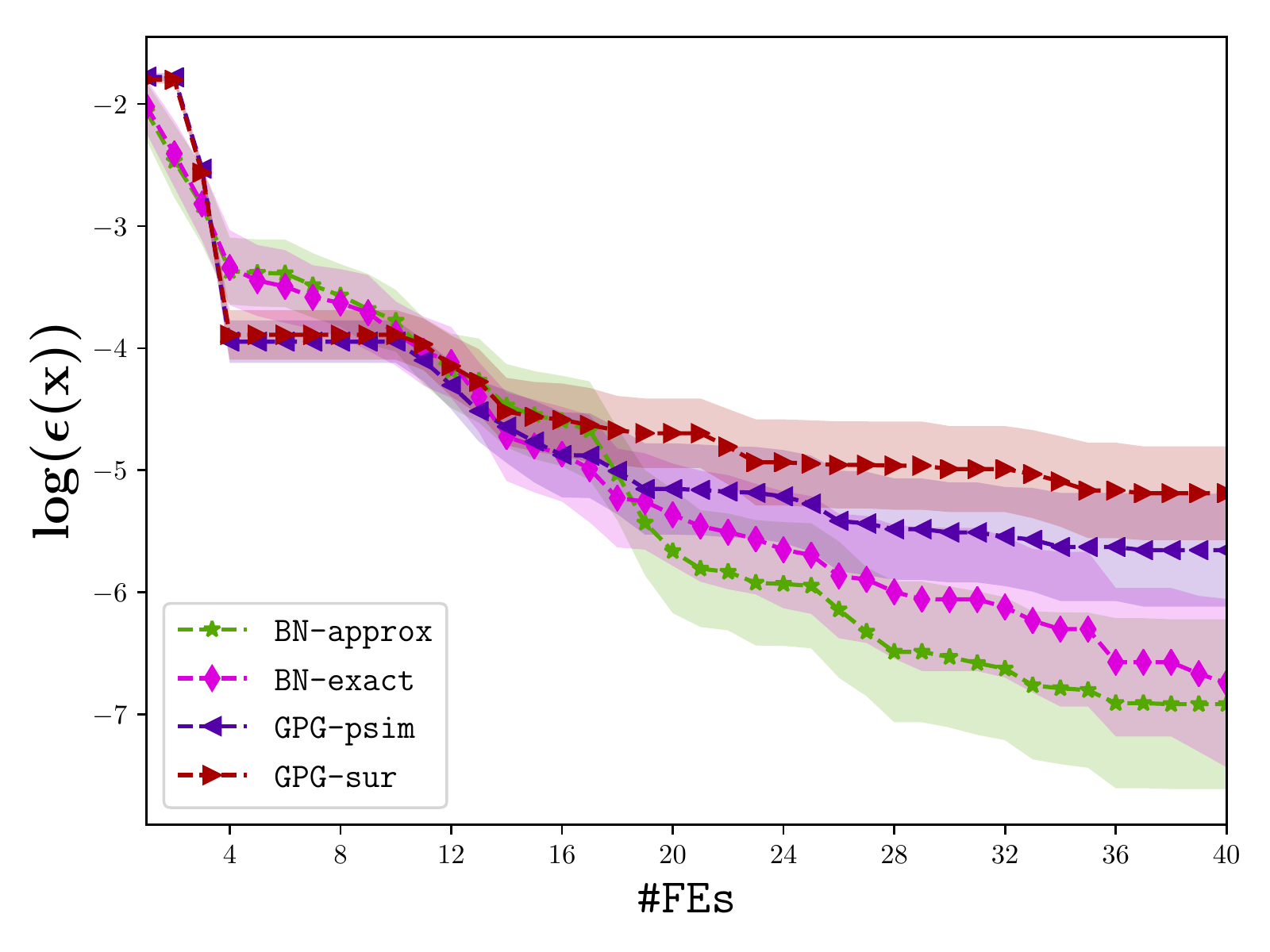}&
		\includegraphics[width=0.33\textwidth]{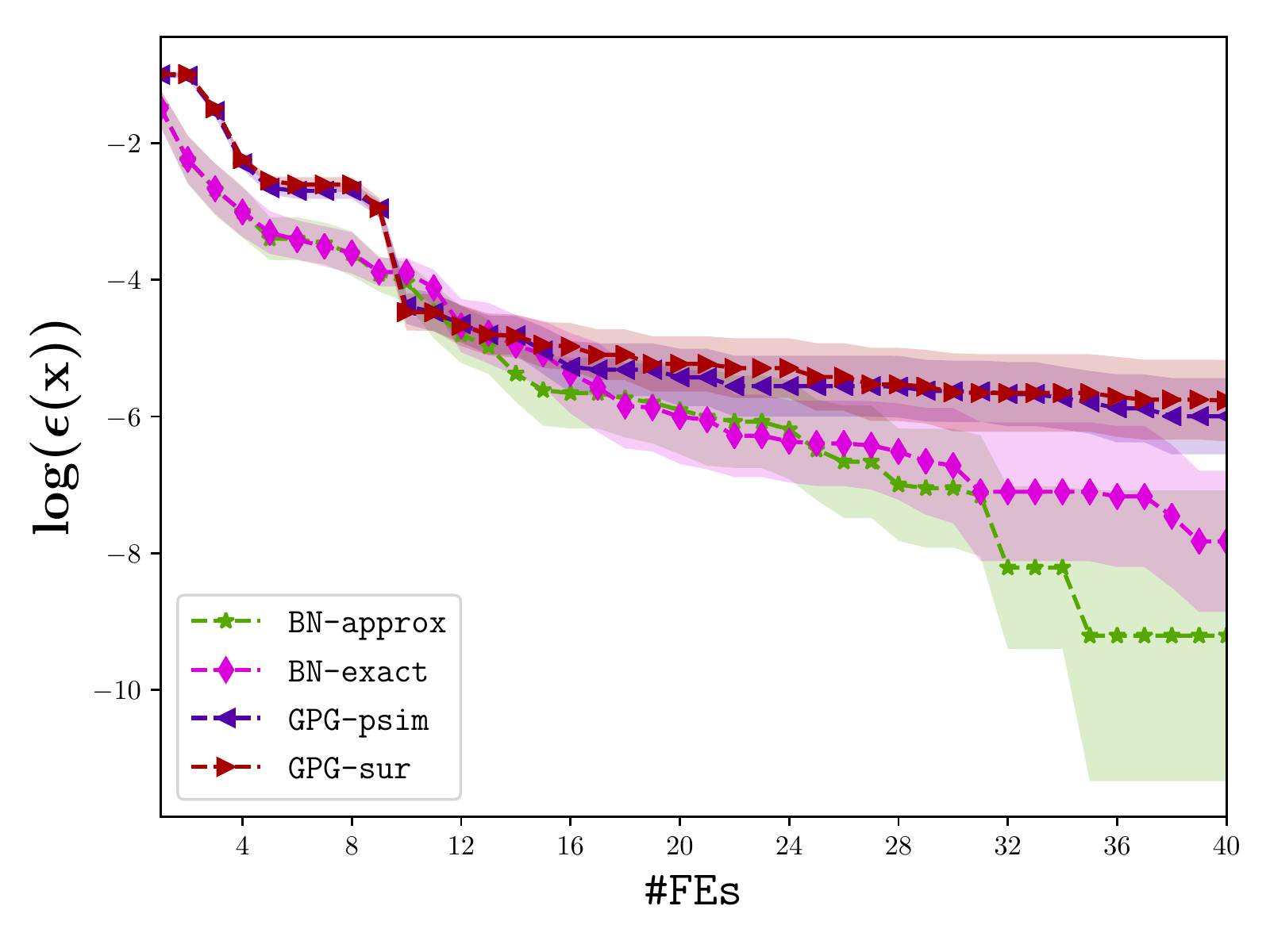}&
		\includegraphics[width=0.33\textwidth]{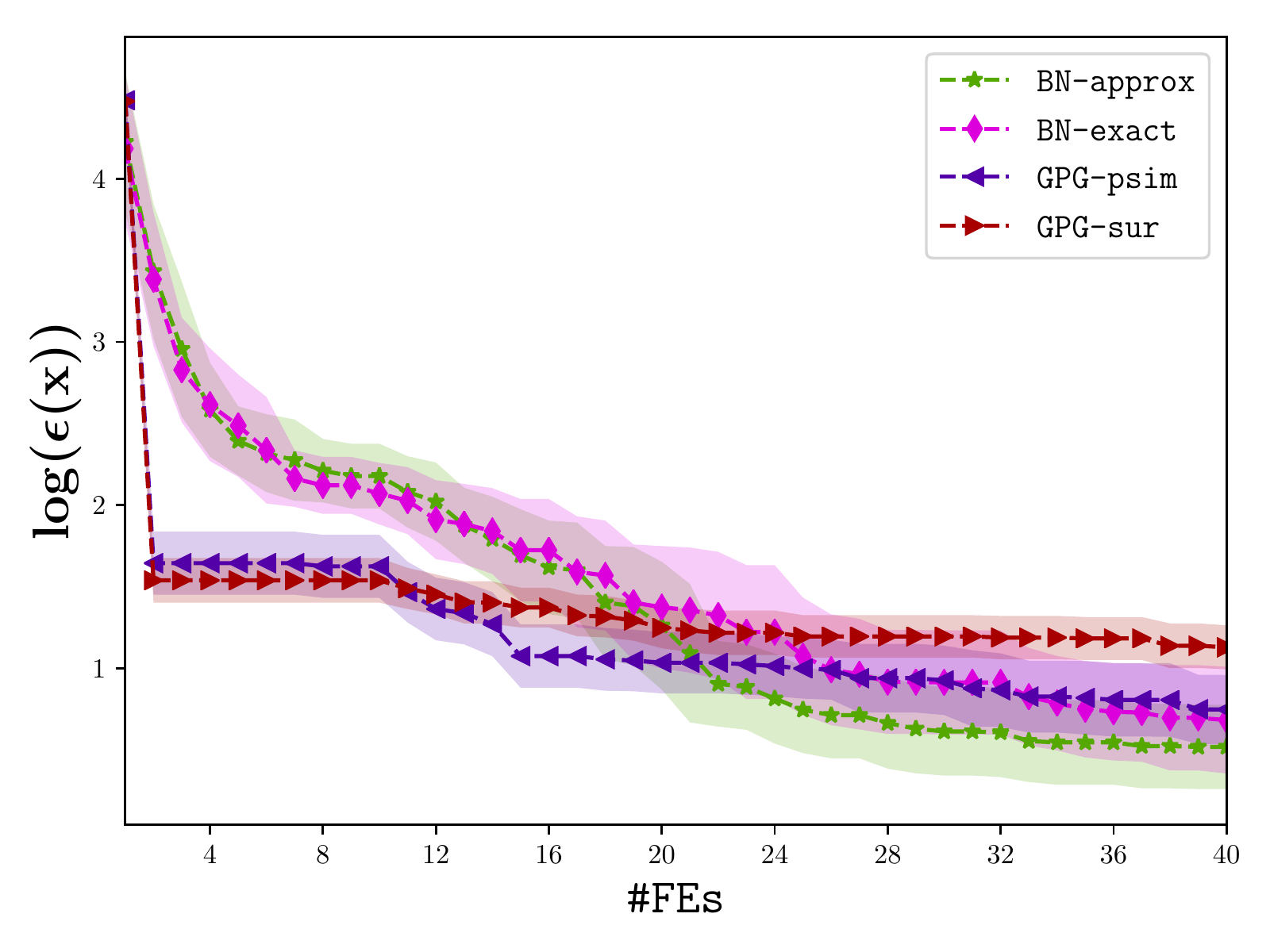} \\
	\end{tabular}
	\caption{Regret $\epsilon(\vx)$ convergence on a log scale as
          a function of the number of function evaluations \fes,
          obtained using $25$ independent runs. The markers indicate
          the average regret value. The error bands corresponds to one
          standard deviation.}
        \label{fig:results_regret}
\end{figure}

\paragraph{Results.} Fig.~\ref{fig:results_regret} shows the convergence of the best obtained regret~$\epsilon$~(\eqref{eq:regret}) as a function of the number of function evaluations over $25$ independent runs. Regret is calculated numerically using \texttt{CMA-ES}. We start with \saddle.1 where the \NE is more appropriately placed for
the grid used by \GPG. In the noiseless version
(Fig.~\ref{fig:results_regret}-a), we observe the lowest regret for the
\GPG variants.  Both \bneq variants have higher regret than the \GPG
variants. Still, the \BR method has the highest regret. In the noisy
\saddle.1 (Fig.~\ref{fig:results_regret}-b), the regret is higher for
all the methods. But \bneq variants show a relatively better performance in comparison to the noiseless version of the problem (Fig.~\ref{fig:results_regret}-a), when
noise is added, the \bneqa has the lowest, then \bneqe, both are lower
than \GPGp and \GPGs. For noiseless \saddle.2 (Fig.~\ref{fig:results_regret}-c) the
NE is shifted off the grid that \GPG uses. Here, we observe the
lowest regret for the \bneq variants. Both \bneq variants have lower
regret than the \GPG variants. The \BR method has the highest
regret. In the noisy \saddle.2 (Fig.~\ref{fig:results_regret}-d) all
methods have higher regret. Again, the order of the algorithms final
regret when noise is added, the \bneqa has the lowest, then \bneqe,
both are lower than \GPGp and \GPGs. The effect of discretizing
$\calX$ shows up clearly in the inferior performance of \GPG variants over the
same problem but with translated (shifted) NEs, namely \saddle.1 and
\saddle.2. 

For noiseless \mop (Fig.~\ref{fig:results_regret}-e) we observe the
lowest regret for the \bneq variants. The \bneqa has the lowest
average regret in the end. Both \bneq variants have lower regret than
the \GPG variants. The \BR method has the highest regret. When noise
is added (Fig.~\ref{fig:results_regret}-f), all methods
have higher regret. There are some changes in the ranking of the
algorithms based on final regret, the \bneqa has the lowest, then
\bneqe, is slightly lower than \GPGp, and the highest regret is
reported for \GPGs.

\begin{figure}[t!]
  \centering
  \includegraphics[width=0.35\textwidth]{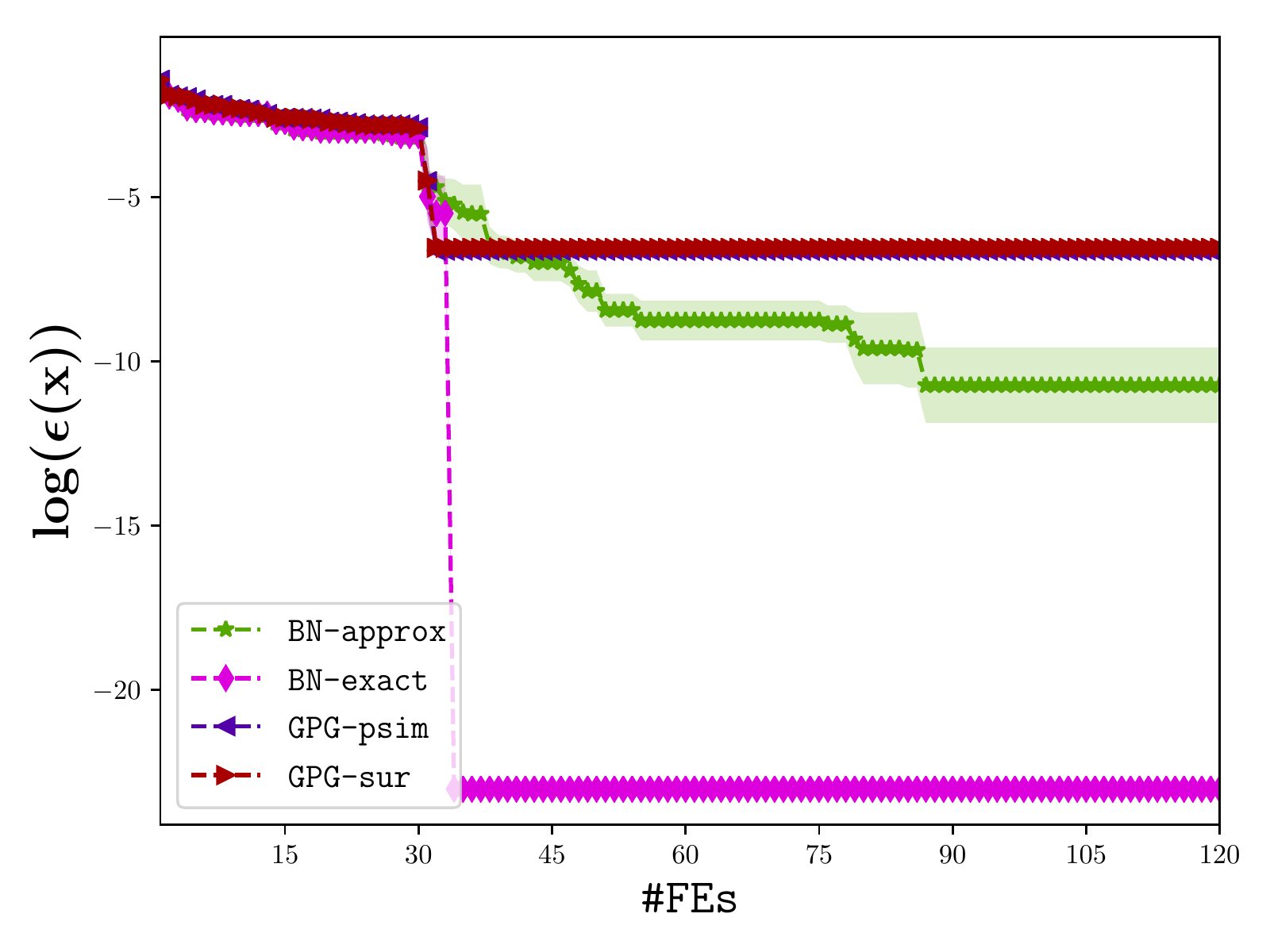}
  \caption{Regret $\epsilon(\vx)$ convergence on a log scale as a
    function of the number of function evaluations \fes, obtained
    using $8$ independent runs. The markers indicate the average
    regret value. The error bands corresponds to one standard
    deviation.}
        \label{fig:results_regret_long}
\end{figure}

We scaled the easy (for \GPG) variant of \saddle~to $4$ dimensions ( \saddle.3) and studied its performance when the number of function evaluations are almost equal to the number
of \GPG's grid points. As $31\times 31$ function evaluations were computationally prohibitive, we used a grid size of $11 \times 11$ for \GPG.  From Fig.~\ref{fig:results_regret_long}, we observe
the lowest regret for the \bneq. The \GPG versions are not capable on
improving the regret, because they are limited to the grid. However the \bneq variants, as the number of \texttt{FEs} increases, surpass the minimum regret of \GPG.
From the results, we see that \GPG performs well and rapidly finds
solutions with low regret when the grid provides the appropriate bias
for the search. The \GPG starts to struggle when the \NE is shifted or
the discretization is too wide. This is the gap in performance that
the \bneq variants address. The continuous representation of \bneq
means that it is not limited to the grid and can continue to improve
the regret with more fitness evaluations. A drawback of the \bneq
representation is that it can take longer to reach low regret. This tradeoff is acceptable because it is searching the full space and not constrained to only find estimates that occupy the grid. Another benefit of
\bneq is that the memory usage is not dependent on discretization. Setting \GPG's grid size to $93 \times 93$ produced a memory allocation error on a $24$-core Ubuntu machine with a $24$-GB RAM.  The tradeoff between \bneq and \GPG is the available \fes, when~$\fes \leq 25$, \GPG methods are in general preferable.

\section{Conclusion}
\label{sec:conclusion}

In this paper, we presented \bneq: a \texttt{B}ayesian framework for
computing \texttt{N}ash equilibria for expensive, black-box,
continuous games. In contrast to proposed methods in the literature,
which either solve a bi-level optimization problem or assume a
discretized representation of the strategy space, our framework seeks
to jointly learn the full game and estimate the game-theoretic regret
on the \emph{full} continuous strategy space. The approach was validated on a
collection of synthetic games and compared to existing methods. We
show that \bneq is capable of improving the regret in noisy
and high-dimensional games to an extent which hierarchical or discretized
methods are not able to reach in an expensive setup. The experiments demonstrated \bneq's
robustness to noise and translated NEs. Secondarily, we observed
that the randomness introduced by the empirical computation of the
approximated regret (\bneqa) can be helpful in directing the search,
compared to computing it exactly (\bneqe). This is in line with the
exploration-exploitation dilemma, in the sense that noisy approximated
regret can contribute to the exploration component of our search for
NE. Future work will include exploring the parameters of \bneq, incorporating inseparable kernels, and applying
it to a simulation-based model for a Cybersecurity application.

\bibliographystyle{splncs04}
\bibliography{bibliography}
\newpage

\pagebreak
\section*{
	\centering Supplementary Materials for: \\
Approximating Nash Equilibria for Black-Box Games: A Bayesian Optimization Approach}

\begin{subappendices}
	\renewcommand{\thesection}{\Alph{section}}%
	% or try \arabic{section}

In this extra material, we derive the closed-form expression for our approximation~(\eqref{eq:regret-approx}) of the regret~(\eqref{eq:regret}). First, the notations are tabulated for readability. Second, the closed form of $\bar{\mu}_i(\vx|\dset)$~(\eqref{eq:barmu}) and $\bar{\sigma}_i(\vx | \dset)$~(\eqref{eq:barstd}) are derived. Besides exact computations, these quantities can be computed empirically as mentioned in~Section~\ref{sec:methods} of the main paper. Finally, we demonstrate our proposition with both \emph{exact} and \emph{approximate} computation  on the \saddle1 and \mop~problems.

\section{Table of Notations}

\begin{table}
	\centering
		\caption{Table of Notations.}
	\begin{tabular}{ll}
		\toprule
		\textbf{Notation} & \textbf{Definition} \\ \toprule 
		$I$ & Set of players \\
		$p$ & Number of players, $|I|=p$ \\
		$\calXi$ & Strategy (action) space of player $i$ \\
		$\calXni$ & Joint strategy (action) space of all players except player $i$\\
		$\calX$ & Joint strategy (action) space of all players. That is, $\calX=\prod_{i=1}^{p}\calXi$ \\
		$n_{\calXi}$ & Number of strategy variables of player $i$. That is, $\calXi \subseteq \R^{n_{\calXi}}$ \\
		$n_{\calX}$ & Number of joint strategy variables. That is, $n_\calX=\sum_{i\in I} n_\calXi$ \\
		$\vxi$ & Action of player $i$. That is, $\vxi\in \calXi$\\
		$\vx$ & Joint actions of all players~$I$ \\
		$\epsilon(\vx)$ & Regret (maximum benefit from unilateral gain) \\
		$\hat{\epsilon}(\vx)$ & Approximated regret according to the GP models \\
		$\I_{\calXi}$  & Subset of coordinate indices of $\calX$ that corresponds to $\calXi$ \\
		\bottomrule 
	\end{tabular}
\end{table}

\newpage 
\section{Analytical Expression for $\bar{\mu}_i(\vx|\dset)$}
\label{sec:mu-proof}
From \eqref{eq:barmu}, we have
\begin{eqnarray}
\muripos{\vx} &=& \E_{\vxi^\prime}[\mu_i(\vxi^\prime, \vxni|\dset)] \nonumber\\
&=& \int_{\calXi} \mufipos{\vxip, \vxni} p(\vxip)d\vxip \nonumber
\\
&=& \int_{\calXi}  \vki(\vxip, \vxni)^T \vKi^{-1}o^{1:t}_i p(\vxip)d\vxip \nonumber
\\
&=& \vqi(\vxni)^T \vKi^{-1}o^{1:t}_i\;,
\end{eqnarray}
where $p(\vxip) = \calU(\vxip \mid \calXi)$ and, without loss of generality, $\calXi=[0,1]^{n_\calXi}$
\begin{eqnarray}
q_{i,j}(\vxni) &=& \int_{\calXi} k_i((\vxip,\vxni), \vx^{(j)})d\vxip \nonumber\\
				&=& v_i\int_{\calXi}  \delta((\vxip, \vxni)-\vx^{(j)}) d\vxip + c_i \nonumber\\ 
				&& \quad \times \int_{\calXi} c_i\exp\bigg(-\frac{((\vxip, \vxni)-\vx^{(j)})^TD_i ((\vxip, \vxni)-\vx^{(j)})}{2}\bigg) d\vxip  \nonumber\\
				&=& v_i \indicator\{\vxni  = \vxni^{(j)}\} + c_i \exp\bigg(-\frac{(\vxni - \vxni^{(j)})^TD_i(\text{-}i,\text{-}i)(\vxni - \vxni^{(j)})}{2}\bigg)\nonumber \\ 
				&& \quad \quad\quad \times \int_{\calXi} \exp\bigg(-\frac{(\vxip - \vxi^{(j)})^TD_i(i,i)(\vxip - \vxi^{(j)})}{2}\bigg) d\vxip \nonumber \\
				&=& v_i \indicator\{\vxni  = \vxni^{(j)}\} + c_i \exp\bigg(-\frac{(\vxni - \vxni^{(j)})^TD_i(\text{-}i,\text{-}i)(\vxni - \vxni^{(j)})}{2}\bigg)\nonumber \\
				&&  \times \; \prod_{l\in \I_{\calXi}} 
				\sqrt{\frac{\pi}{2D^{l,l}_i}} 
				\Bigg(\erf\Bigg(x_{l}^{(j)}\sqrt{\frac{D^{l,l}_i}{2}}\Bigg) + 
				\erf\Bigg((1 - x_{l}^{(j)})\sqrt{\frac{D^{l,l}_i}{2}}\Bigg) \Bigg) \nonumber\;,
\end{eqnarray}
where $D_i(i,i)$ denotes the submatrix that lies in the rows and columns of $D_i$ that correspond to the action space (coordinates) of player $i$, $\I_{\calXi}$ is the subset of coordinate indices of $\calX$ that corresponds to $\calXi$'s coordinates. $D^{l,l}_i$ denotes the $(l,l)$ entry of matrix $D_i$ and $x_{l}^{(j)}$ is the value of $l^{th}$ coordinate in $\calX$ of $j^{th}$ strategic play $\vx^{(j)}\in \dset$ and $1 \leq j \leq t$.

\begin{comment}
\begin{eqnarray}
\kfipos{\vx}{\vx^{\prime}} = k_i(\vx, \vx^{\prime}) - \vki(\vx)^T\vKi^{-1}\vki(\vx^{\prime})
\end{eqnarray}

With $p(\vxni, \vxni^\prime)=p(\vxni)p(\vxni^\prime)= \calU(\vxni, \vxni^\prime \mid \calXni^2)$
\begin{eqnarray}
\kripos{\vxi}{\vxi^{\prime}} &=& \int_{\calXni} \int_{\calXni^\prime} \kfipos{\vx}{\vx^{\prime}} p(\vxni, \vxni^\prime) d\vxni d\vxni^\prime\nonumber \\
&=& \int_{\calXni} \int_{\calXni^\prime} k_i(\vx, \vx^{\prime}) p(\vxni, \vxni^\prime) d\vxni d\vxni^\prime \\ && \quad - \quad \int_{\calXni} \int_{\calXni^\prime} \vki(\vx)^T\vKi^{-1}\vki(\vx^{\prime}) p(\vxni, \vxni^\prime) d\vxni d\vxni^\prime \nonumber \\
&=& \int_{\calXni} \int_{\calXni^\prime} k_i((\vxi, \vxni), (\vxi^{\prime}, \vxni^{\prime})) d\vxni d\vxni^\prime  \nonumber \\ 
&& \quad - \quad \int_{\calXni} \int_{\calXni^\prime} \vki((\vxi, \vxni))^T\vKi^{-1}\vki((\vxi^{\prime}, \vxni^{\prime})) p(\vxni, \vxni^\prime) d\vxni d\vxni^\prime \nonumber \\
&=&  \int_{\calXni^\prime} \underline{k}_i(\vx, (\vxi^\prime, \vxni^\prime)) p(\vxni^\prime)d\vxni^\prime -  \vqi(\vxi)^T \vKi^{-1}\vqi(\vxi^\prime) \\
&=& v_i \indicator\{\vxi  = \vxi^\prime\} - \vqi(\vxi)^T \vKi^{-1}\vqi(\vxi^\prime)   + c_i \exp\bigg(-\frac{(\vxi - \vxi^{\prime})^TD_i(i,i)(\vxi - \vxi^{\prime})}{2}\bigg)\nonumber \\
&& \quad \quad \times \quad {\prod}^{n_{\calXni}}_{l=1}  \frac{1}{d_i(l,l)}\bigg(\sqrt{2\pi  d_i(l,l)} \erf\bigg(\sqrt{\frac{d_i(l,l)}{2}}\bigg) + 2 \exp\bigg(-\frac{d_i(l,l)}{2}\bigg) -2\bigg) \nonumber \\
&&  \nonumber 
\end{eqnarray}

\end{comment}
\newpage 
\section{Analytical Expression for $\bar{\sigma}_i(\vx | \dset)$}
\label{sec:std-proof}

From~\eqref{eq:barstd}, we have
\begin{eqnarray}
\bar{\sigma}^2_i(\vx | \dset)&=& \E_{\vxi^\prime}[(\mu_i(\vxi^\prime, \vxni|\dset)- \bar{\mu}_i(\vx|\dset))^2] \nonumber \\
&=& \E_{\vxi^\prime}[(\mu_i(\vxi^\prime, \vxni|\dset))^2]- \bar{\mu}_i(\vx|\dset)^2\nonumber\;,
\end{eqnarray}
where $\bar{\mu}_i(\vx|\dset)$ is defined in Appendix~\ref{sec:mu-proof} and
\begin{eqnarray}
\E_{\vxip}[\mufipos{\vxi^\prime, \vxni}^2] &=& \int_{\calXi} (\vki(\vxip,\vxni)^T \vKi^{-1}o^{1:t}_i)^2 p(\vxip) d\vxip \nonumber \\
							&=& \int_{\calXi} (o^{1:t}_i)^T \vKi^{-T} \vki(\vxip,\vxni) \vki(\vxip,\vxni)^T \vKi^{-1}o^{1:t}_i \;d\vxip \nonumber \\
							&=&  (o^{1:t}_i)^T \vKi^{-T} \vQi(\vxni) \vKi^{-1}o^{1:t}_i\;,  \label{eq:var-first}
\end{eqnarray}
where $p(\vxip) = \calU(\vxip \mid \calXi)$ and, without loss of generality, $\calXi=[0,1]^{n_\calXi}$. The $(p,q)$ entry of matrix $\vQi(\vxni)$ is defined in the next page.
\section{Further Illustrations}
\label{sec:fig-illust}

\begin{figure}
	\centering
	\begin{tabular}{cc}
		\includegraphics[width=0.45\textwidth]{figs/saddle_dec_space_bayes_5iters.pdf} &
		\includegraphics[width=0.45\textwidth]{figs/saddle_dec_space_bayes_20iters.pdf} \\
		\tiny (a) Sampled profiles after $5$ observations & \tiny (b) Sampled 
		profiles after $20$ observations\\
		\includegraphics[width=0.5\textwidth,clip,trim=62 100 50 100]{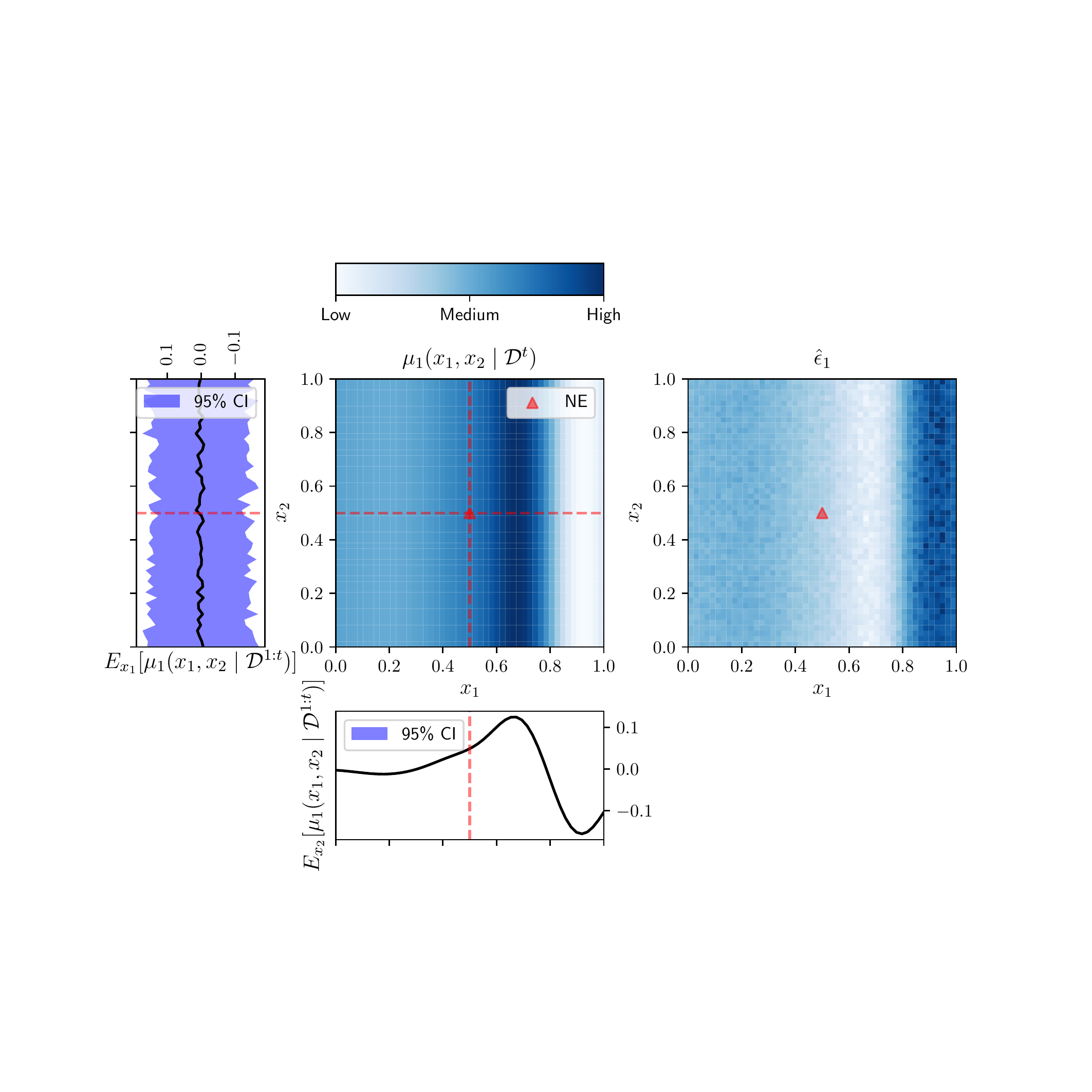} &
		\includegraphics[width=0.5\textwidth,clip,trim=50 100 62 100]{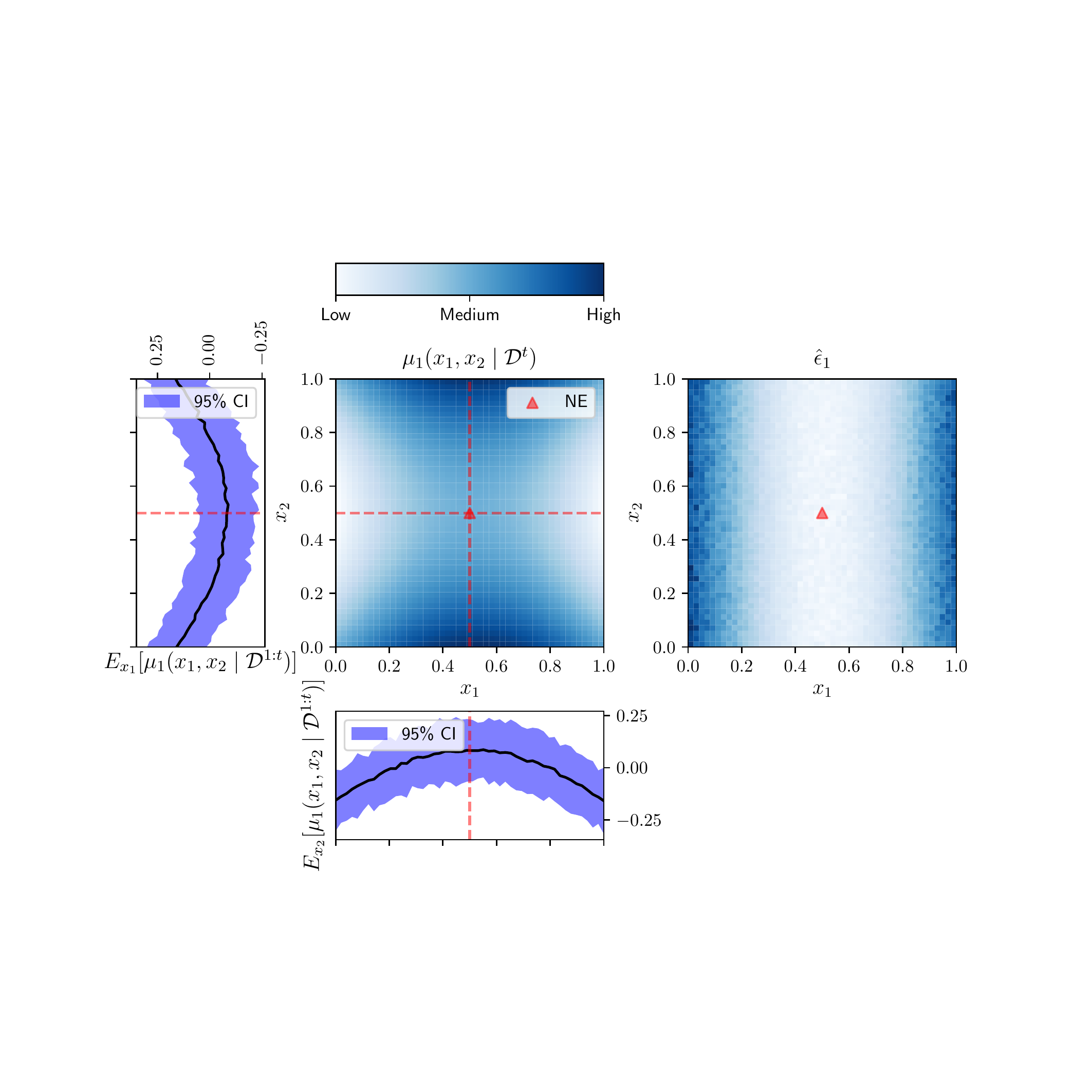} \\
		\tiny (c) Player $1$  after $5$ observations&
		\tiny (d) Player $1$ after $20$ observations \\
		\includegraphics[width=0.5\textwidth,clip,trim=62 100 50 100]{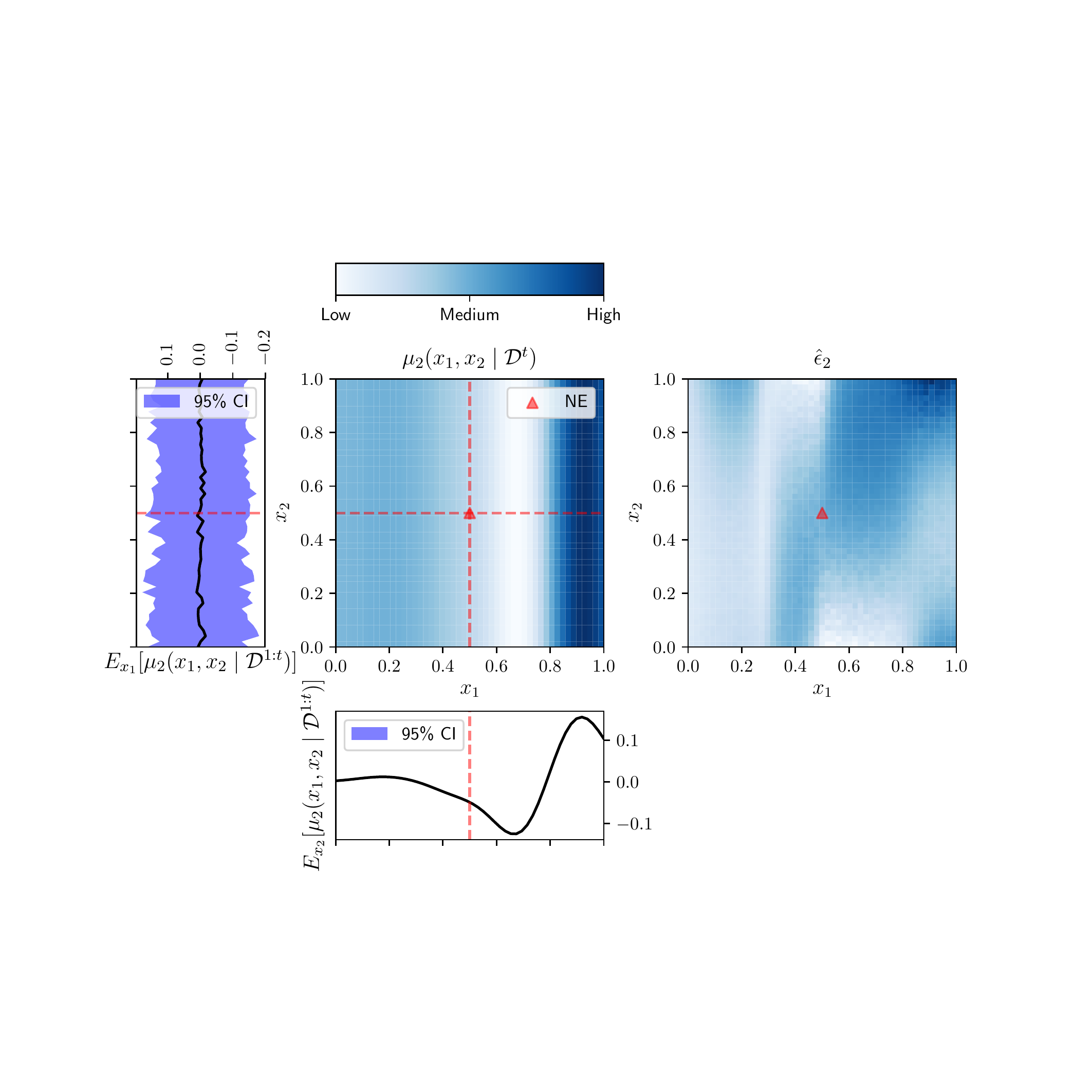} &
		\includegraphics[width=0.5\textwidth,clip,trim=50 100 62 100]{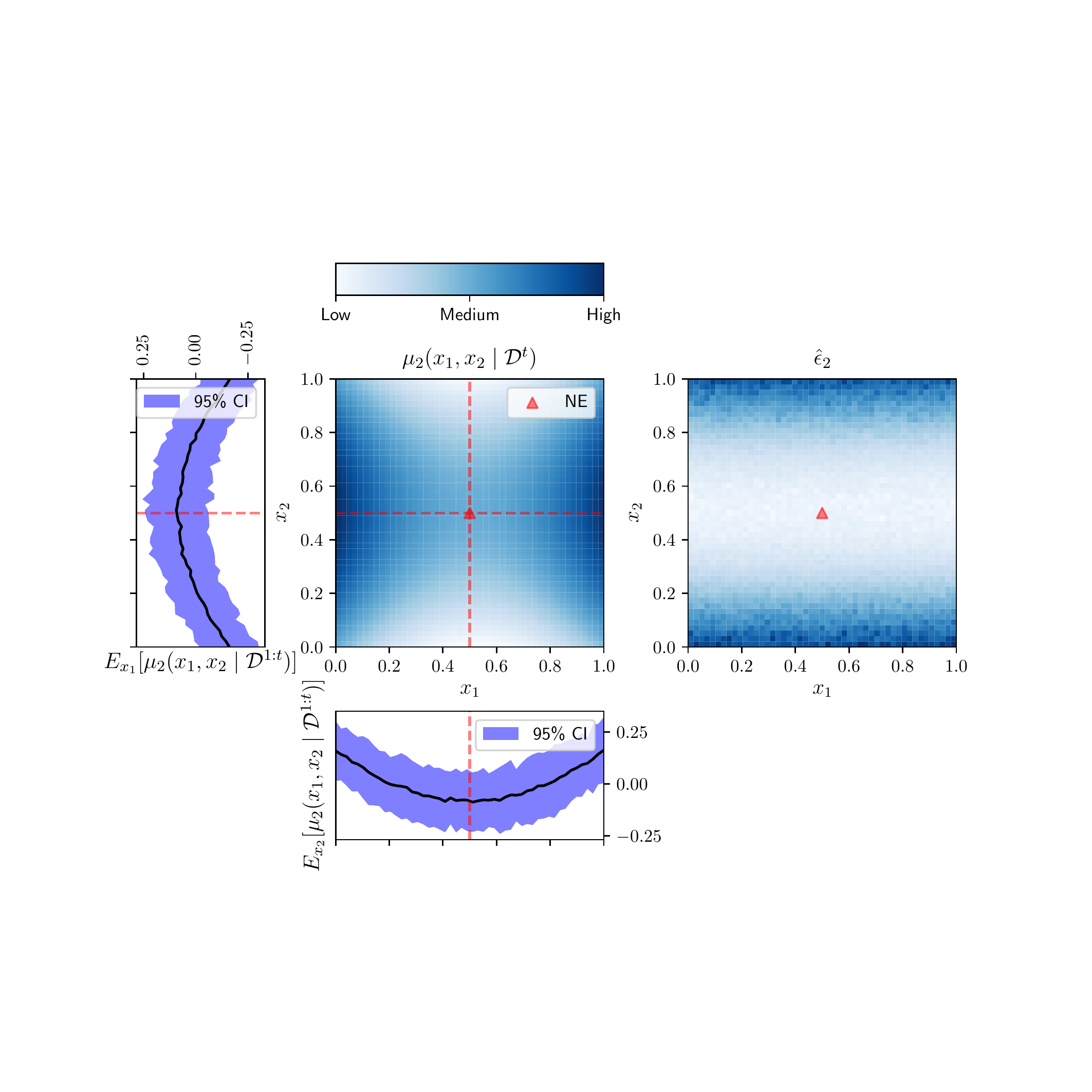} \\
		\tiny (e) Player $2$ after $5$ observations & \tiny
		(f) Player $2$ after $20$ observations \\
	\end{tabular}
	\caption{Illustration of \bneqa after $5$ and $20$ observations on the zero-sum hyperbolic paraboloid game with NE being $(0.5,0.5)$. That is, $u_1(x_1, x_2)= (x_2 - 0.5)^2 - (x_1 - 0.5)^2$, and $u_2(x_1, x_2)=-u_1(x_1, x_2)$. $\hat{\epsilon}_i$ denotes $(\hat{\bar{\mu}}_i(\vx|\dset) + \gamma \hat{\bar{\sigma}}_i(\vx | \dset)  -  \mu_i(\vx|\dset))/ \hat{\bar{\sigma}}_i(\vx | \dset)$, where $\gamma=2.32635$, the $99^{th}$ percentile of the standard normal distribution.}
	\label{fig:saddle-illust-approx}
\end{figure}

\begin{figure}
	\centering
	\begin{tabular}{cc}
		\includegraphics[width=0.45\textwidth]{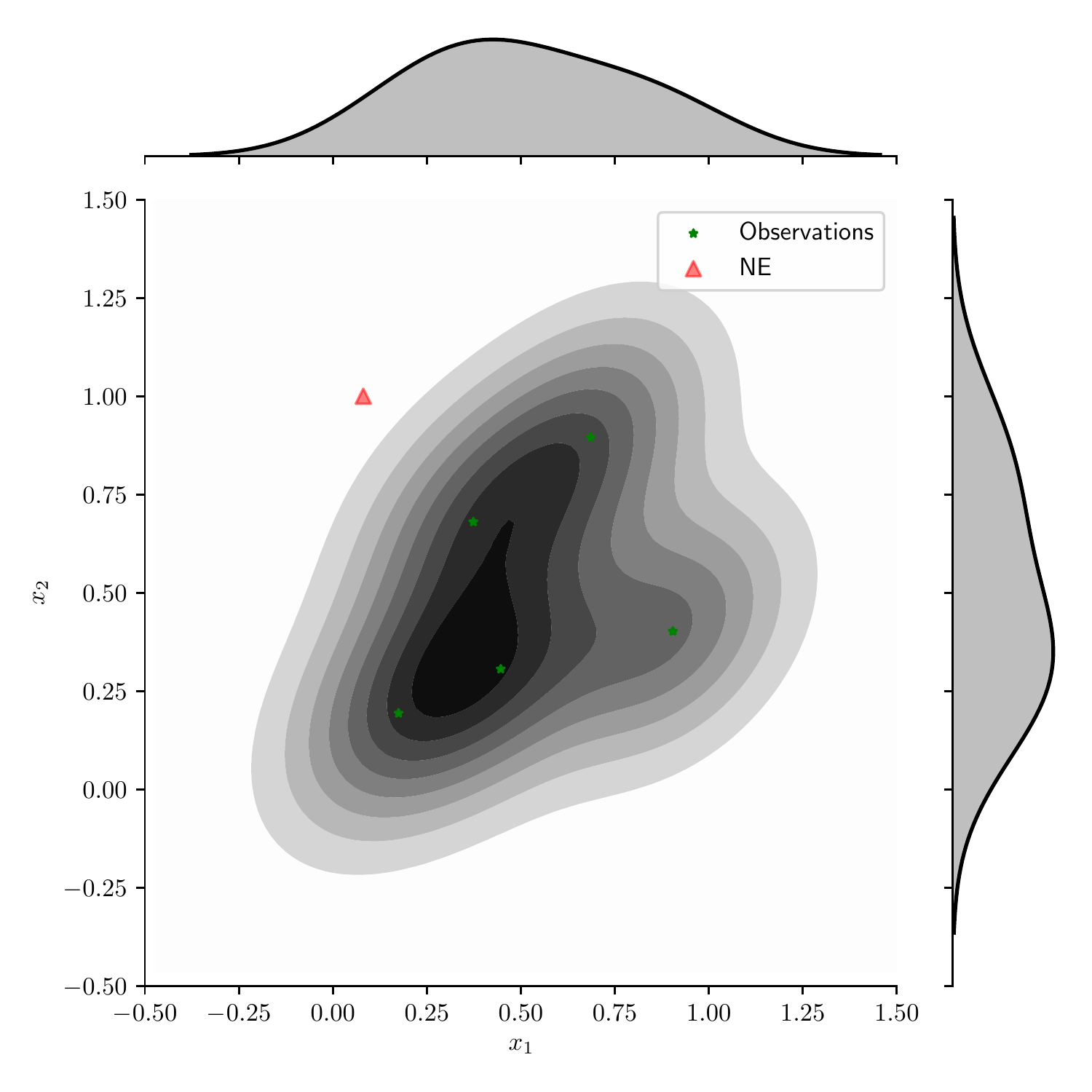} &
		\includegraphics[width=0.45\textwidth]{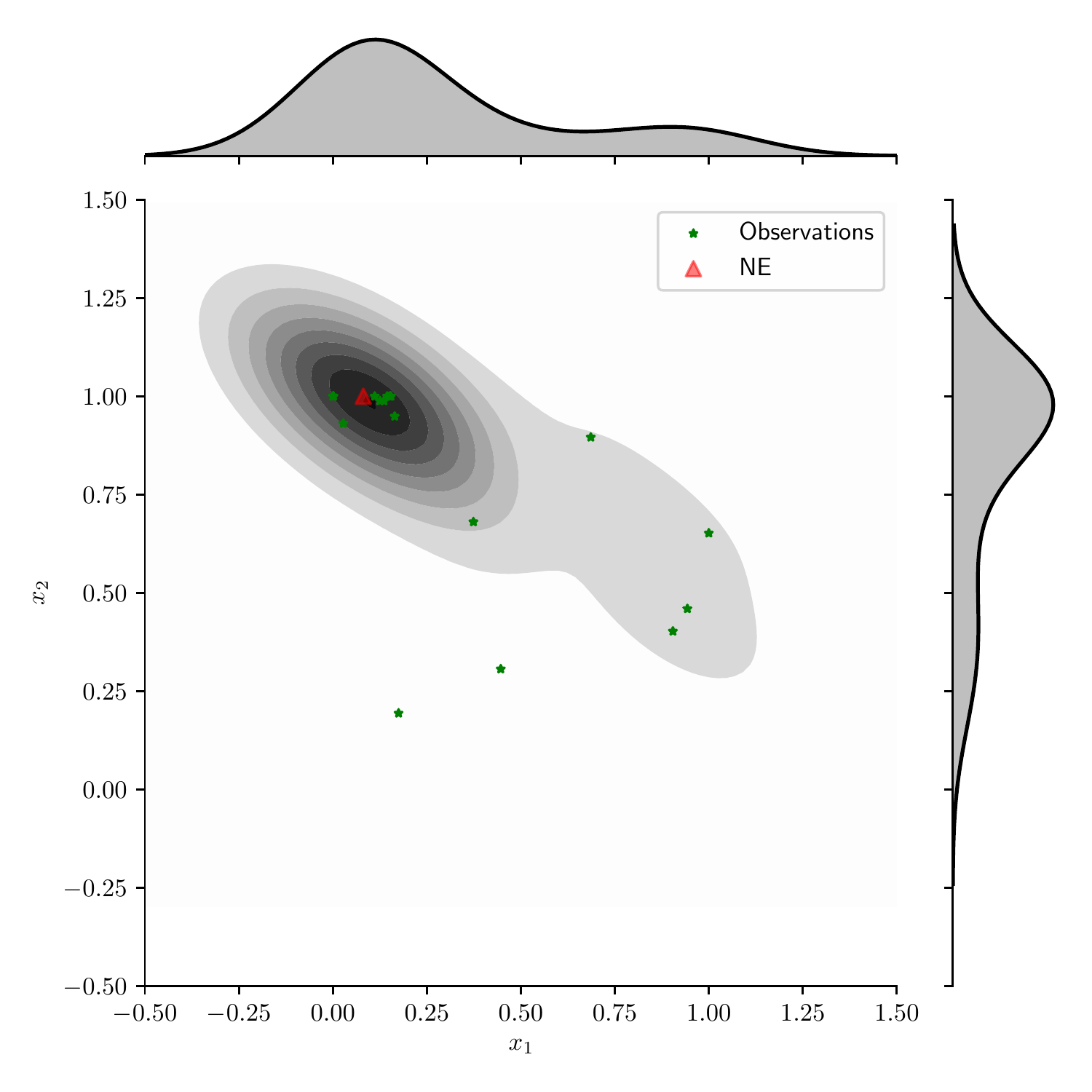} \\
		\tiny (a) Sampled profiles after $5$ observations & \tiny (b) Sampled profiles $\vx$ after $20$ observations\\
		\includegraphics[width=0.5\textwidth,clip,trim=62 100 50 100]{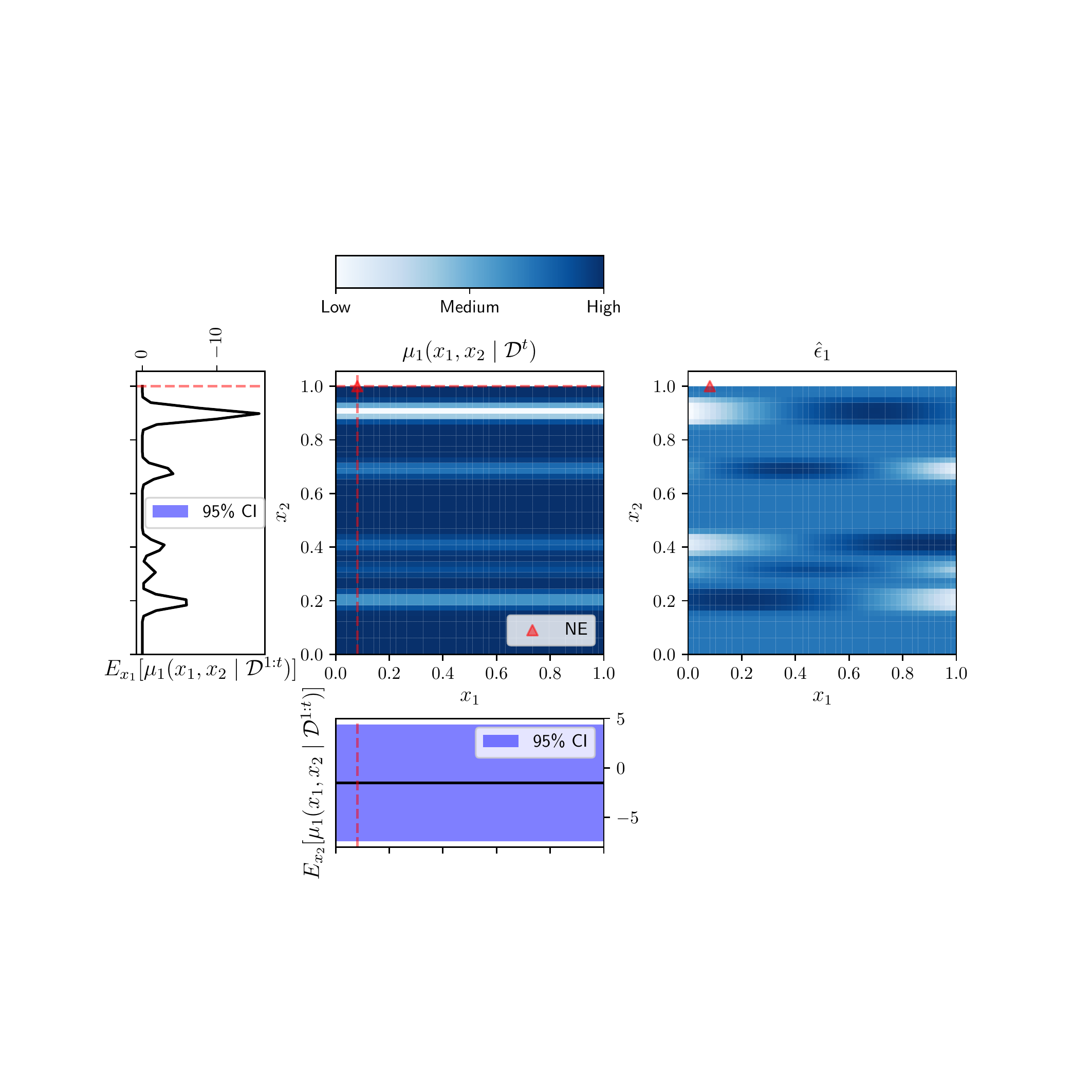} &
		\includegraphics[width=0.5\textwidth,clip,trim=50 100 62 100]{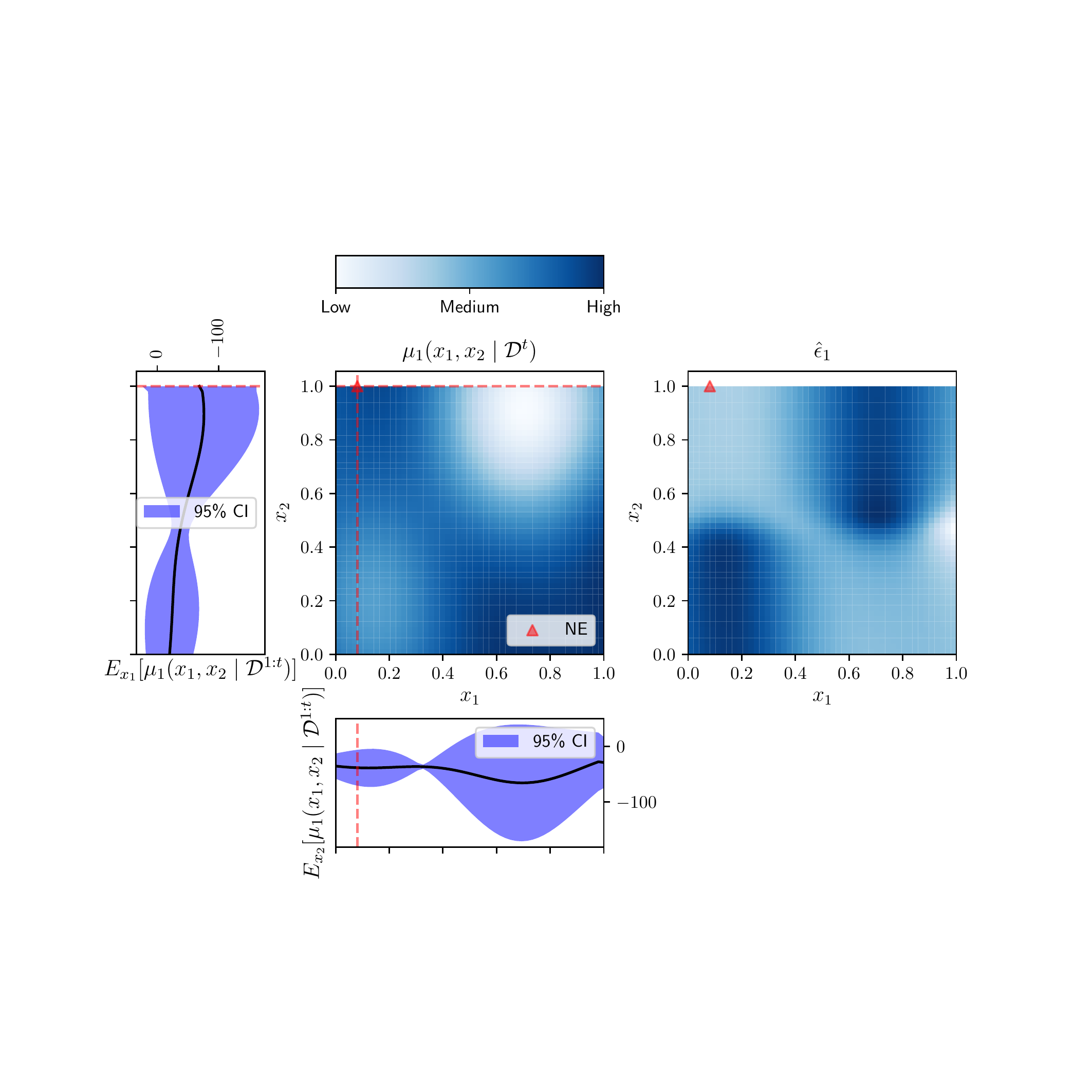} \\
		\tiny (c) Player $1$ after $5$ observations &
		\tiny (d) Player $1$  after $20$ observations \\
		\includegraphics[width=0.5\textwidth,clip,trim=62 100 50 100]{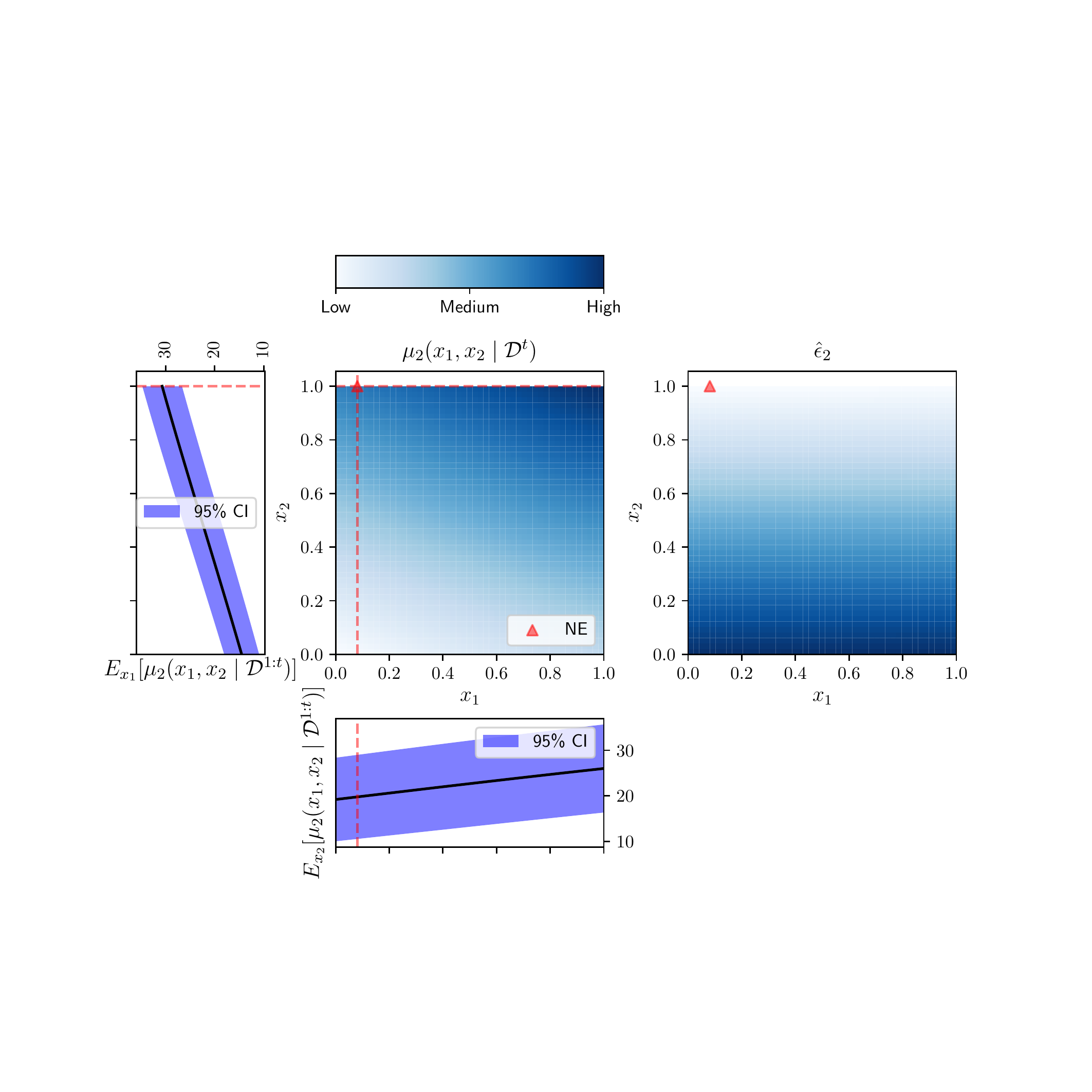} &
		\includegraphics[width=0.5\textwidth,clip,trim=50 100 62 100]{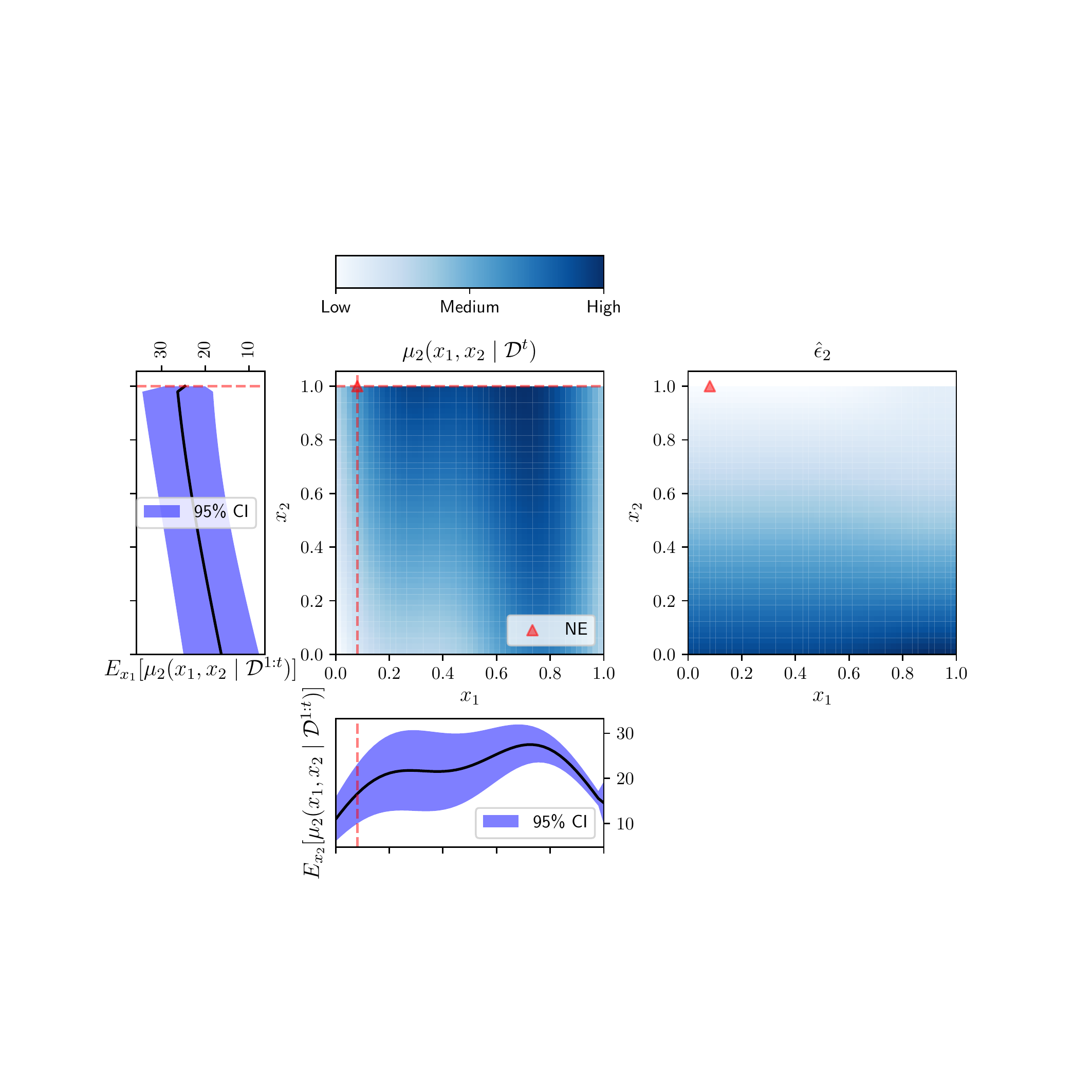} \\
		\tiny (e) Player $2$  after $5$ observations & \tiny
		(f) Player $2$ after $20$ observations \\
	\end{tabular}
	\caption{Illustration of \bneqe after $5$ and $20$ observations on the classical multi-objective problem considered in~\cite{picheny2016bayesian}. $\hat{\epsilon}_i$ denotes $(\bar{\mu}_i(\vx|\dset) + \gamma \bar{\sigma}_i(\vx | \dset)  -  \mu_i(\vx|\dset))/ \bar{\sigma}_i(\vx | \dset)$, where $\gamma=2.32635$, the $99^{th}$ percentile of the standard normal distribution.}
	\label{fig:mop-illust-exact}.
\end{figure}

\begin{figure}
	\centering
	\begin{tabular}{cc}
		\includegraphics[width=0.45\textwidth]{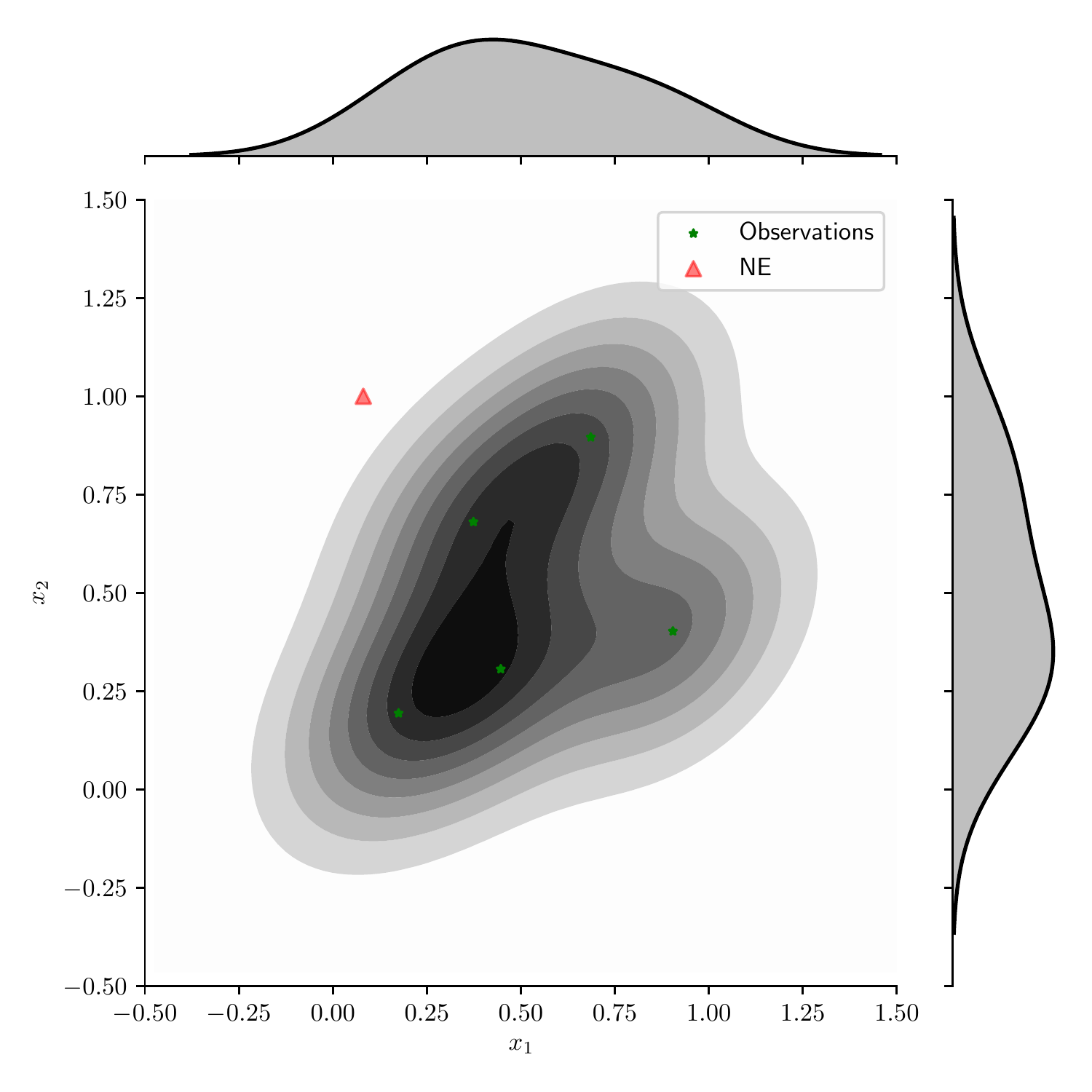} &
		\includegraphics[width=0.45\textwidth]{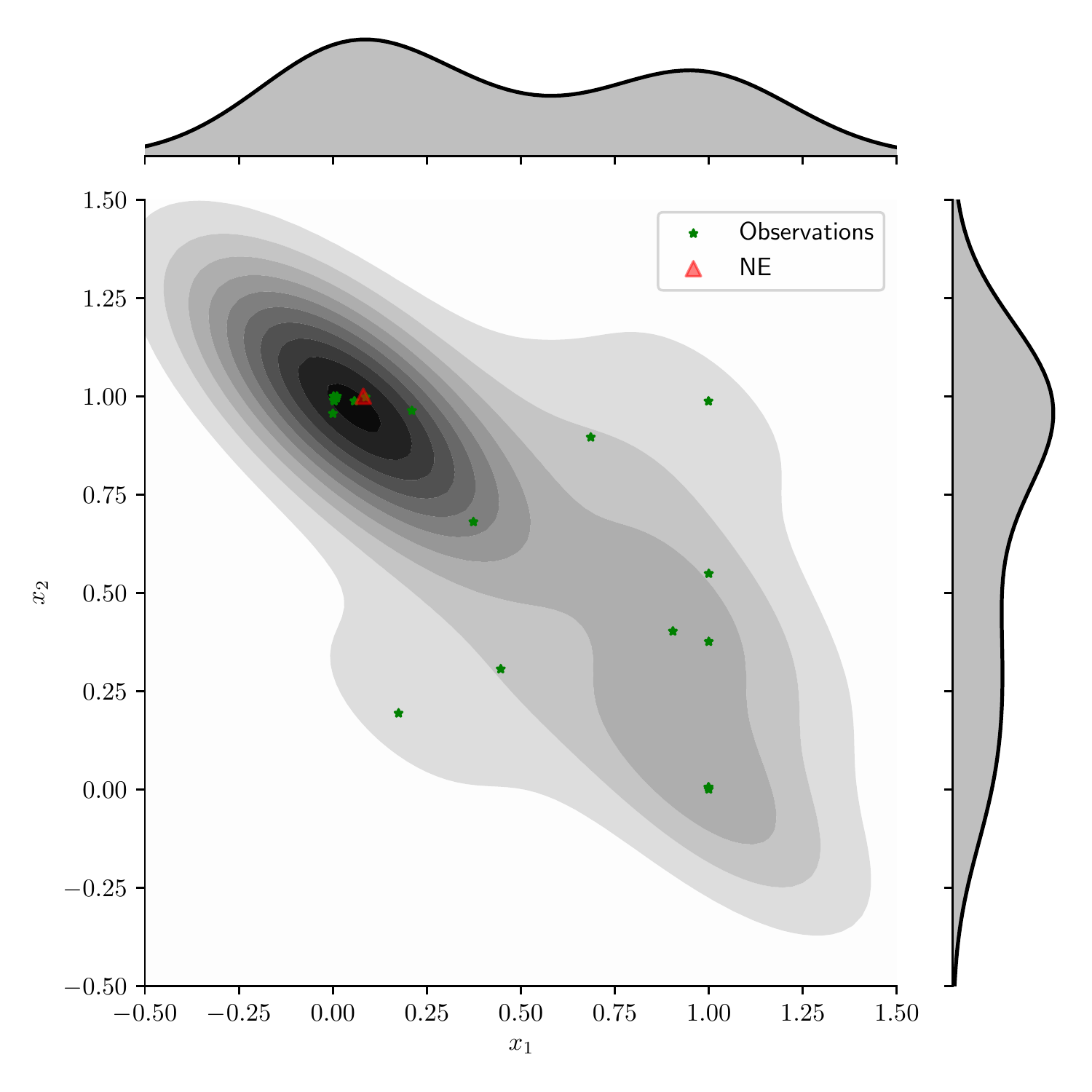} \\
		\tiny (a) Sampled profiles after $5$ observations & \tiny (b) Sampled profiles $\vx$ after $20$ observations\\
		\includegraphics[width=0.5\textwidth,clip,trim=62 100 50 100]{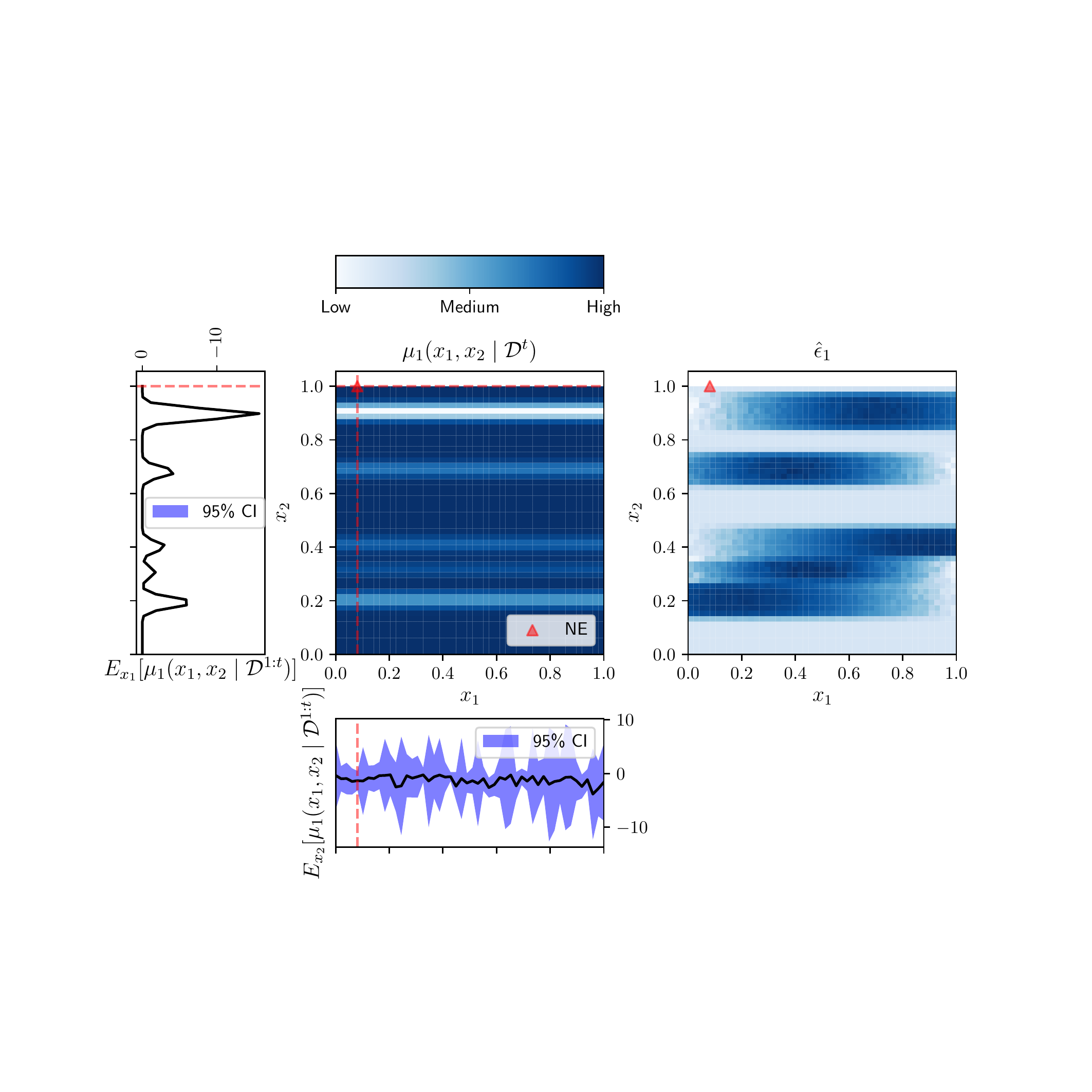} &
		\includegraphics[width=0.5\textwidth,clip,trim=50 100 62 100]{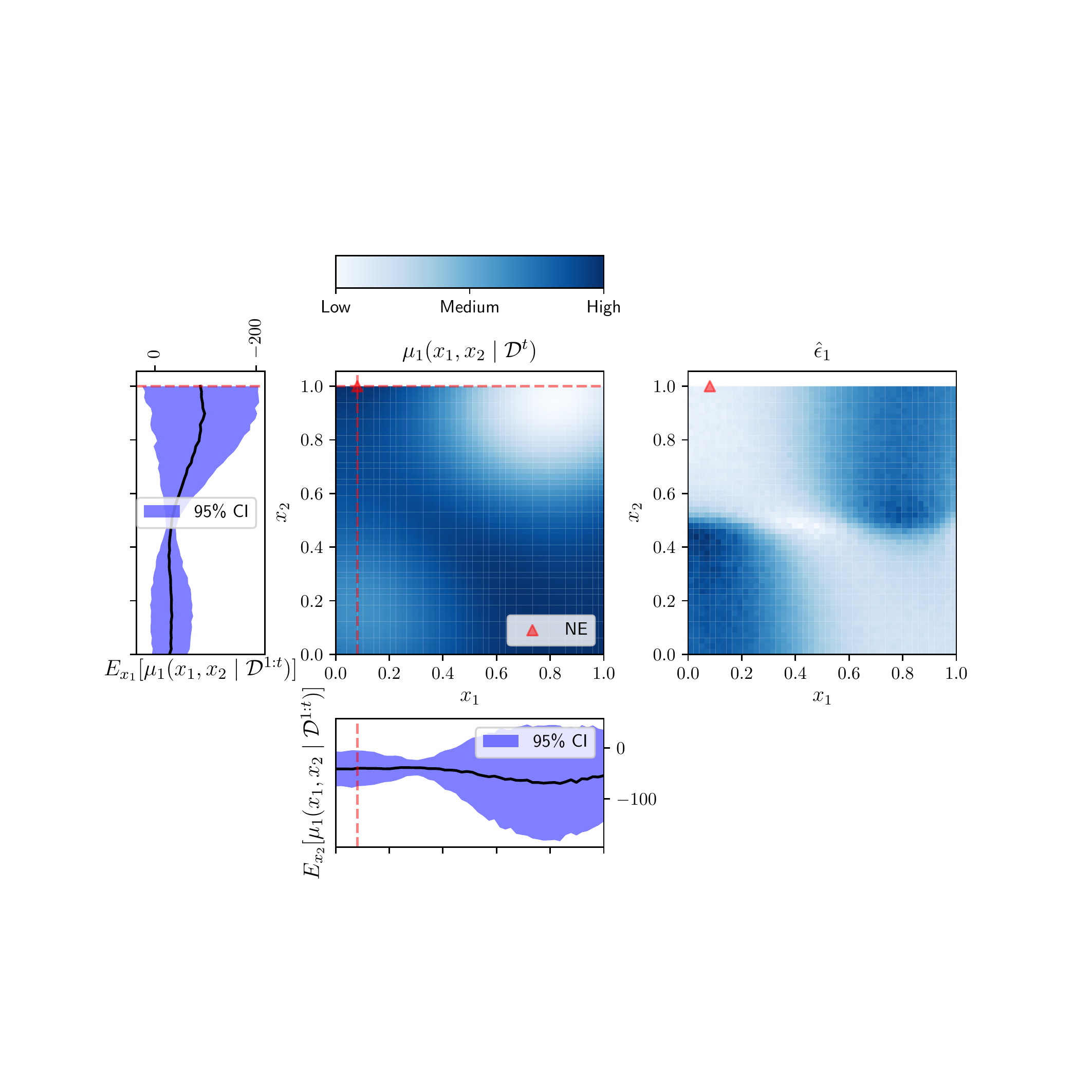} \\
		\tiny (c) Player $1$  after $5$ observations &
		\tiny (d) Player $1$  after $20$ observations \\
		\includegraphics[width=0.5\textwidth,clip,trim=62 100 50 100]{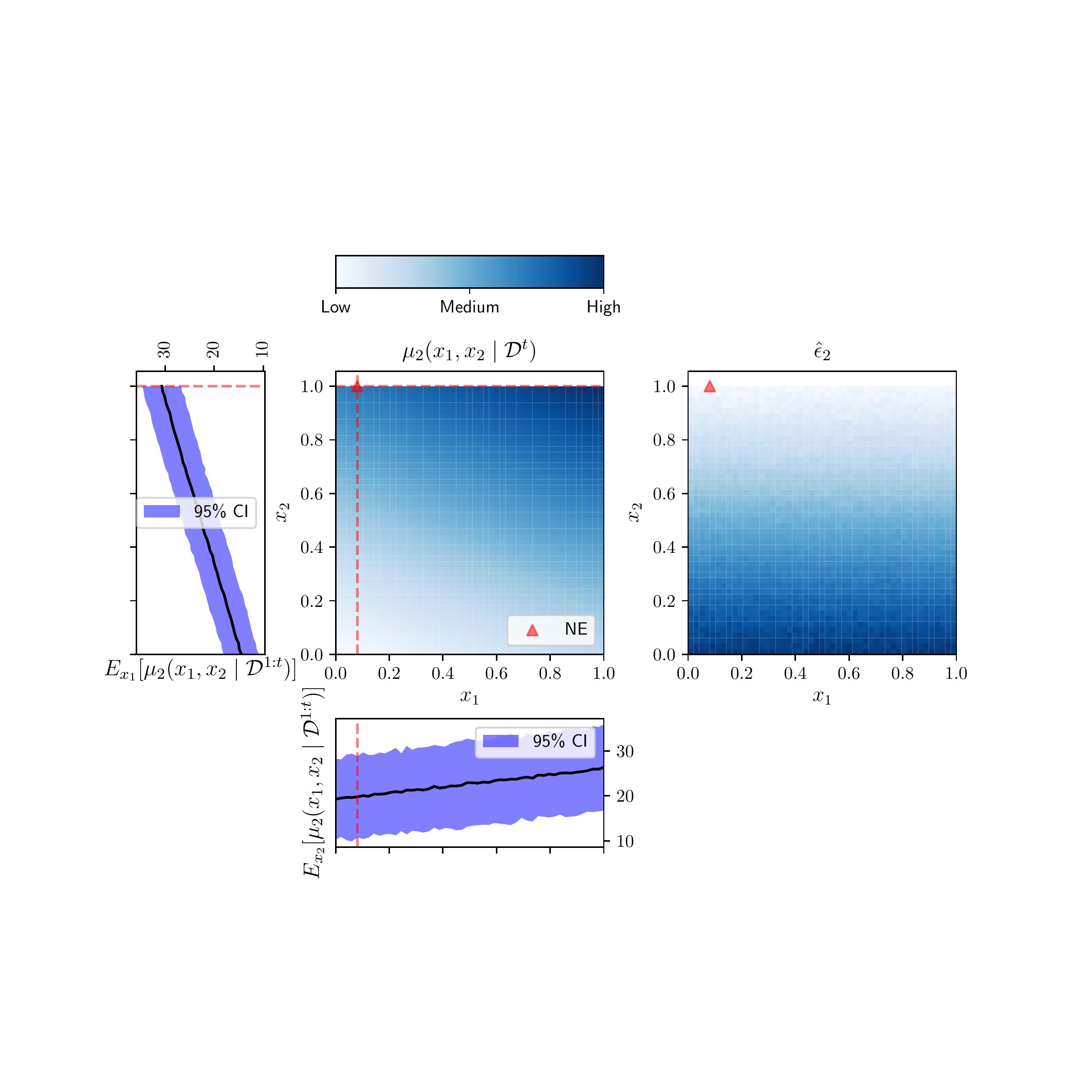} &
		\includegraphics[width=0.5\textwidth,clip,trim=50 100 62 100]{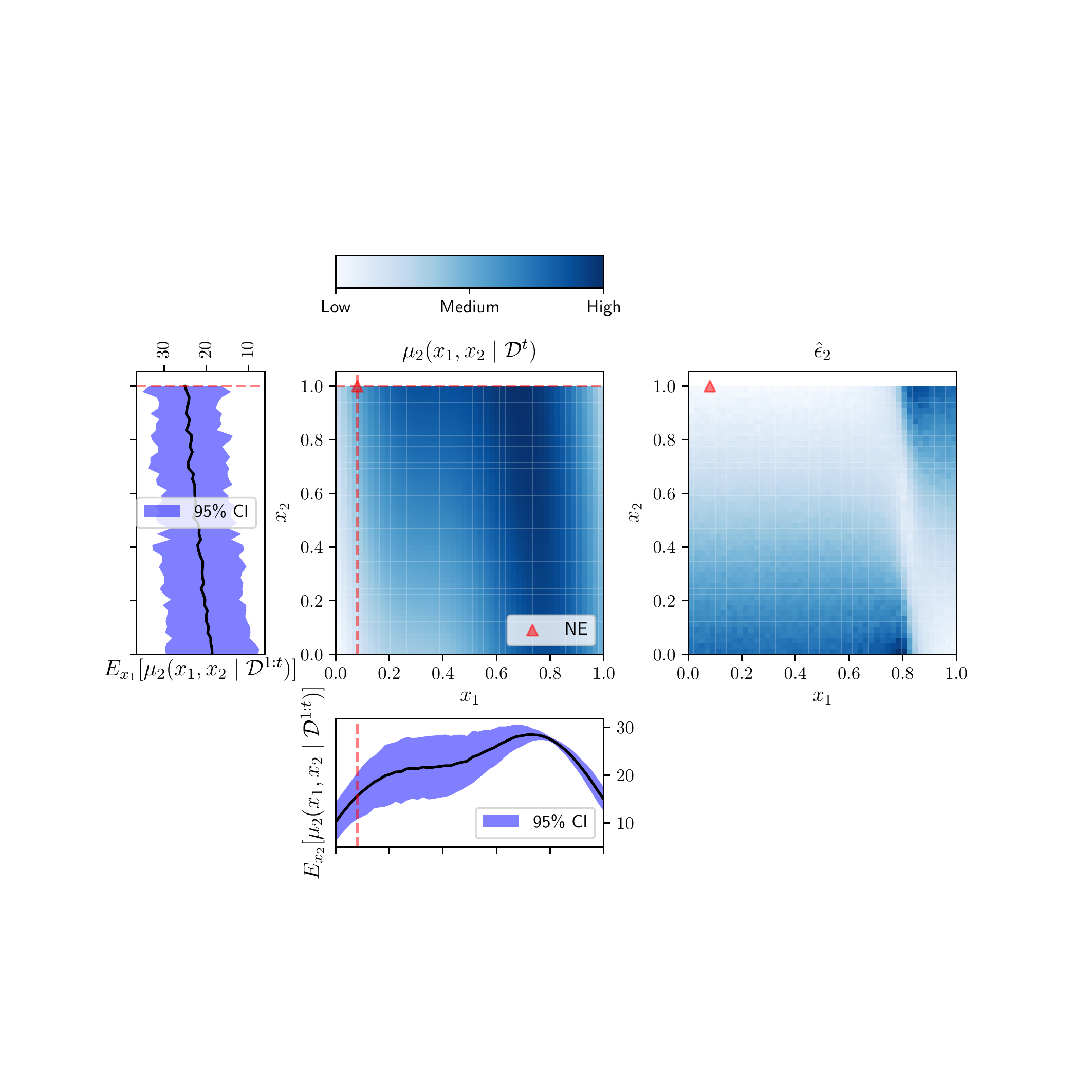} \\
		\tiny (e) Player $2$ after $5$ observations & \tiny
		(f) Player $2$ after $20$ observations \\
	\end{tabular}
	\caption{Illustration of \bneqa after $5$ and $20$ observations on the classical multi-objective problem considered in~\cite{picheny2016bayesian}. $\hat{\epsilon}_i$ denotes $(\hat{\bar{\mu}}_i(\vx|\dset) + \gamma \hat{\bar{\sigma}}_i(\vx | \dset)  -  \mu_i(\vx|\dset))/ \hat{\bar{\sigma}}_i(\vx | \dset)$, where $\gamma=2.32635$, the $99^{th}$ percentile of the standard normal distribution.}
	\label{fig:mop-illust-approx}
\end{figure}

In this section, we complement Fig.~\ref{fig:saddle-illust-exact} with further illustrations of the algorithm in its \emph{exact} and \emph{approximate} variants on the \saddle1 and \mop~problems as shown in Fig.~\ref{fig:saddle-illust-approx}, Fig.~\ref{fig:mop-illust-exact}, and Fig.~\ref{fig:mop-illust-approx}. The figures show the progress of \bneq variants from $5$ to $20$ observations. The top subfigures (a and b) show the sampled profiles and their kernel density estimate (KDE) over the action space~$\calX$. With more iterations, the sampled profiles get closer to the actual NE. The middle subfigures (c and d) show the estimated payoff and approximated regret for player~$1$ after $5$ and $20$ observations, respectively. Likewise, the bottom subfgures (e and f) show the same for player~2. Since subfigures (c), (d), (e), and (f) are similar. We describe them collectively. The center heatmap represents the GP's current estimate of player~$i$'s payoff $\mu_i(x_1, x_2|\calD^{1:t})$~(\eqref{eq:mu}). The supplots to the left and bottom show the expectation (along with a 95\%-confidence interval) of $\mu_i(x_1, x_2|\calD^{1:t})$ over $x_1$ and $x_2$, respectively. For player $1$, the \emph{left} supplot of subfigures (c) and (d)  
represent ${\bar{\mu}}_1(\vx|\calD^{1:5})$~(\eqref{eq:barmu}) and ${\bar{\mu}}_1(\vx|\calD^{1:20})$ in Fig.~\ref{fig:saddle-illust-exact} and~\ref{fig:mop-illust-exact}. Whereas in Fig.~\ref{fig:saddle-illust-approx} and~\ref{fig:mop-illust-approx}, they correspond to $\hat{\bar{\mu}}_1(\vx|\calD^{1:5})$~(\eqref{eq:bbarmu}) and $\hat{\bar{\mu}}_1(\vx|\calD^{1:20})$. For player $2$, the \emph{bottom} subplots in subfigures (e) and (f) correspond to ${\bar{\mu}}_2(\vx|\calD^{1:5})$ and ${\bar{\mu}}_2(\vx|\calD^{1:20})$ in Fig.~\ref{fig:saddle-illust-exact} and~\ref{fig:mop-illust-exact}; and $\hat{\bar{\mu}}_2(\vx|\calD^{1:5})$ and $\hat{\bar{\mu}}_2(\vx|\calD^{1:20})$ in Fig.~\ref{fig:saddle-illust-approx} and~\ref{fig:mop-illust-approx}. The right subplot of the subfigures represents the scaled regret of player~$i$: $(\bar{\mu}_i(\vx|\dset) + \gamma \bar{\sigma}_i(\vx | \dset)  -  \mu_i(\vx|\dset))/\bar{\sigma}_i(\vx | \dset)$ in Fig.~\ref{fig:saddle-illust-exact} and~\ref{fig:mop-illust-exact}; and
$(\hat{\bar{\mu}}_i(\vx|\dset) + \gamma \hat{\bar{\sigma}}_i(\vx | \dset)  -  \mu_i(\vx|\dset))/\hat{\bar{\sigma}}_i(\vx | \dset)$ in Fig.~\ref{fig:saddle-illust-approx} and~\ref{fig:mop-illust-approx}. For player~$1$ (i.e., subfigures (c) and (d)), this is computed based on the center and the left subplots. For player~$2$ (i.e., subfigures (e) and (f)), this is computed based on the center and the \emph{right} subplots. One can observe the smoothness of the plots in the figures of \bneqe (Fig.~\ref{fig:saddle-illust-exact} and~\ref{fig:mop-illust-exact}), in comparison to those of the approximate variant (Fig.~\ref{fig:saddle-illust-approx} and~\ref{fig:mop-illust-approx}).

\begin{eqnarray}
Q^{p,q}_i(\vxni) &=& \int_{\calXi} k_i((\vxip,\vxni), \vx^{(p)}) k_i((\vxip,\vxni), \vx^{(q)}) p(\vxip) d\vxip \nonumber \\
	&=& \int_{\calXi} \Bigg( v_i \delta((\vxip,\vxni)-\vx^{(p)}) +  c_i\exp\bigg(-\frac{((\vxip,\vxni)-\vx^{(p)})^TD_i ((\vxip,\vxni)-\vx^{(p)})}{2}\bigg) \Bigg) \nonumber \\
	&& \times \Bigg(v_i \delta((\vxip,\vxni)-\vx^{(q)}) +  c_i\exp\bigg(-\frac{((\vxip,\vxni)-\vx^{(q)})^TD_i ((\vxip,\vxni)-\vx^{(q)})}{2}\bigg)\Bigg)  d\vxip \nonumber \\
	&=& \int_{\calXi} \Bigg( v^2_i \delta((\vxip,\vxni)-\vx^{(p)}) \delta((\vxip,\vxni)-\vx^{(q)}) \nonumber \\
	&& \quad + v_i c_i \delta((\vxip,\vxni)-\vx^{(p)})\exp\bigg(-\frac{((\vxip,\vxni)-\vx^{(q)})^TD_i ((\vxip,\vxni)-\vx^{(q)})}{2}\bigg) \nonumber \\
	&& \quad + v_i c_i \delta((\vxip,\vxni)-\vx^{(q)})\exp\bigg(-\frac{((\vxip,\vxni)-\vx^{(p)})^TD_i ((\vxip,\vxni)-\vx^{(p)})}{2}\bigg) \nonumber \\ 
	&& \quad + c^2_i \exp\bigg(-\frac{((\vxip,\vxni)-\vx^{(p)})^TD_i ((\vxip,\vxni)-\vx^{(p)})}{2}\bigg) \nonumber \\
	&& \quad \quad \times \exp\bigg(-\frac{((\vxip,\vxni)-\vx^{(q)})^TD_i ((\vxip,\vxni)-\vx^{(q)})}{2}\bigg)
	\Bigg)d\vxip \nonumber \\
	&=& v^2_i \indicator\{\vxni  = \vxni^{(p)}, \vx^{(p)}= \vx^{(q)}\} \nonumber \\
	&& \quad  + v_i c_i \indicator\{\vxni  = \vxni^{(p)}\}
	\exp\bigg(-\frac{(\vx^{(p)}-\vx^{(q)})^TD_i (\vx^{(p)}-\vx^{(q)})}{2}\bigg) \nonumber \\
	&& \quad +  v_i c_i \indicator\{\vxni  = \vxni^{(q)}\}
	\exp\bigg(-\frac{(\vx^{(q)}-\vx^{(p)})^TD_i (\vx^{(q)}-\vx^{(p)})}{2}\bigg) \nonumber \\
	&& \quad + c^2_i \times \exp\bigg(-\frac{(\vxni - \vxni^{(p)})^TD_i(\text{-}i,\text{-}i)(\vxni - \vxni^{(p)})}{2}\bigg) \nonumber \\
	&& \quad \quad  \times \exp\bigg(-\frac{(\vxni - \vxni^{(q)})^TD_i(\text{-}i,\text{-}i)(\vxni - \vxni^{(q)})}{2}\bigg) \nonumber \\
	&& \quad \quad \times {\prod}_{l\in \I_{\calXi}} \sqrt{\frac{\pi}{4 D^{l,l}_i}} \exp\bigg(-\frac{D^{l,l}_i (x^{(p)}_{l}-x^{(q)}_{l})^2}{4}\bigg) \nonumber \\
	&& \quad \quad  \times \Bigg(\erf\Bigg(\frac{\sqrt{D^{l,l}_i}}{2}(x^{(p)}_{l} + x^{(q)}_{l})\Bigg)  - \erf\Bigg(\frac{\sqrt{D^{l,l}_i}}{2}(x^{(p)}_{l} + x^{(q)}_{l}-2)\Bigg)\Bigg)\;,\label{eq:var-q}
\end{eqnarray}
where $D_i(i,i)$ denotes the submatrix that lies in the rows and columns of $D_i$ that correspond to the action space (coordinates) of player $i$, $\I_{\calXi}$ is the subset of coordinate indices of $\calX$ that corresponds to $\calXi$'s coordinates. $D^{l,l}_i$ denotes the $(l,l)$ entry of matrix $D_i$ and $x_{l}^{(p)}$ is the value of $l^{th}$ coordinate in $\calX$ of $p^{th}$ strategic play $\vx^{(j)}\in \dset$ and $1 \leq p,q\leq t$.

\end{subappendices}

\end{document}